\documentclass{frontiersSCNS} 

\usepackage{url,microtype,subcaption,booktabs}
\newcommand{\cm}{cm$^{-1}$}
\newcommand{\pmol}{P-molecules}
\usepackage{cleveref}
\usepackage{mhchem} 
\usepackage{chemscheme} 
\usepackage{lscape}

\def\keyFont{\fontsize{8}{11}\helveticabold }
\def\firstAuthorLast{Zapata Trujilo {et~al.}} 
\def\Authors{Juan C. Zapata Trujilo$^{1}$, Anna-Maree Syme$^{1}$, Keiran N. Rowell$^{2}$,  Brendan P. Burns$^{3,4}$,  Ebubekir S. Clark$^{1}$, Maire N. Gorman$^{5}$, Lorrie S. D. Jacob$^{1}$,  Panayioti Kapodistrias$^{1}$, David J. Kedziora${^6}$, Felix A. R. Lempriere$^{1}$,Chris Medcraft$^{1}$,  Jensen O'Sullivan$^{1}$, Evan G.  Robertson$^{7}$, Georgia G. Soares$^{4,8}$, Luke Steller$^{4,8}$,  Bronwyn L. Teece$^{4,8}$, Chenoa D. Tremblay$^{9}$,  Clara Sousa-Silva$^{10}$, Laura K. McKemmish$^{1,*}$ }

\def\Address{$^{1}$School of Chemistry, University of New South Wales, Sydney, NSW, 2052, Australia \\
 $^{2}$School of Chemistry, University of Sydney, Sydney, NSW, 2052, Australia \\
 $^{3}$School of Biotechnology and Biomolecular Sciences, University of New South Wales, Sydney, NSW, 2052, Australia \\
 $^4$ Australian Centre for Astrobiology, University of New South Wales, Sydney, NSW, 2052, Australia \\
 $^{5}$Department of Physics, Aberystwyth University, Ceredigion, UK, SY23 3BZ, UK \\
 $^{6}$Advanced Analytics Institute, Faculty of Engineering and Information Technology, University of Technology Sydney, NSW, 2007, Australia \\
 $^{7}$Department of Chemistry and Physics, La Trobe Institute for Molecular Science, La Trobe University, Victoria, 3086, Australia \\

 $^{8}$School of Biological, Earth and Environmental Sciences, University of New South Wales, Sydney, NSW, 2052, Australia \\
 $^{9}$CSIRO Astronomy and Space Science, Bentley WA 6102, Australia\\
 $^{10}$Harvard-Smithsonian Center for Astrophysics, Cambridge, MA, 02138, United States of America.  }

\def\corrAuthor{Laura K. McKemmish}
\def\corrEmail{l.mckemmish@unsw.edu.au}

\newcommand{\package}[1]{#1}
\newcommand{\alert}[1]{\textcolor{black}{ #1}}

\newcommand{\mc}{\multicolumn}
\newcolumntype{H}{>{\setbox0=\hbox\bgroup}c<{\egroup}@{}}

\usepackage{gensymb}

\begin{document}
\onecolumn
\firstpage{1}

\title[Infrared Spectroscopy of Phosphorus-bearing Molecules]{Computational Infrared Spectroscopy of 958 Phosphorus-bearing Molecules} 

\author[\firstAuthorLast ]{\Authors} 
\address{} 
\correspondance{} 

\extraAuth{}

\maketitle

\begin{abstract}

Phosphine is now well established as a biosignature, which has risen to prominence with its recent tentative detection on Venus. To follow up this discovery and related future exoplanet biosignature detections, it is important to spectroscopically detect the presence of phosphorus-bearing atmospheric molecules that could be involved in the chemical networks producing, destroying or reacting with phosphine. 

We start by enumerating phosphorus-bearing molecules (\pmol{}) that could potentially be detected spectroscopically in planetary atmospheres and collecting all available spectral data. Gaseous P-molecules are rare, with speciation information scarce. Very few molecules have high accuracy spectral data from experiment or theory; instead, the best current spectral data was obtained using a high-throughput computational algorithm, \package{RASCALL}, relying on functional group theory to efficiently produce approximate spectral data for arbitrary molecules based on their component functional groups.  

Here, we present a high-throughput approach utilising established computational quantum chemistry methods (CQC) to produce a database of approximate infrared spectra for 958 \pmol{}. These data are of interest for astronomy and astrochemistry (importantly identifying potential ambiguities in molecular assignments), improving \package{RASCALL}'s underlying data, big data spectral analysis and future machine learning applications. However, this data will probably not be sufficiently accurate for secure experimental detections of specific molecules within complex gaseous mixtures in laboratory or astronomy settings. We chose the strongly performing harmonic $\omega$B97X-D/def2-SVPD model chemistry for all molecules and test the more sophisticated and time-consuming GVPT2 anharmonic model chemistry for 250 smaller molecules. Limitations to our automated approach, particularly for the less robust GVPT2 method, are considered along with pathways to future improvements.  

Our CQC calculations significantly improve on existing \package{RASCALL} data by providing quantitative intensities, new  data in the fingerprint region (crucial for molecular identification) and higher frequency regions (overtones, combination bands), and improved data for fundamental transitions based on the specific chemical environment. As the spectroscopy of most \pmol{} has never studied outside \package{RASCALL} and this approach, the new data in this paper is the most accurate spectral data available for most \pmol{} and represent a significant advance in the understanding of the spectroscopic behaviour of these molecules.

\tiny
 \keyFont{\section{Keywords:} infrared spectroscopy, exoplanet atmospheres, phosphine, Venus, phosphorus-bearing molecules, computational quantum chemistry; spectral data; VPT2} 
\end{abstract}

\section{Introduction}

\alert{Phosphine (\ce{PH3}) is currently a strong biosignature candidate as there are few, if any, non-biological formation pathways of phosphine for terrestrial planets \citep{sousa2019phosphine}.} A tentative discovery of phosphine in the cloud decks of Venus was recently reported, with predicted abundances on the order of ppb \citep{20GrRiBa.P}\footnote{We note that the discovery of phosphine on Venus is preliminary. Independent analyses of the data are ongoing and the unambiguous detection of phosphine on Venus will require follow-up observations.} that cannot be explained by non-biological sources \citep{20BaPeSe.P}. To investigate the presence and formation mechanisms of phosphine on Venus, and to interpret future observations of planetary atmospheres, we must improve our understanding of the chemical networks that may include phosphine. A crucial tool in this process is the ability to detect phosphorus-bearing molecules (\pmol{}) that can provide clues to the formation pathways of phosphine, and provide insight into the mechanisms of a possible phosphine-producing biosphere. Gaseous P-molecules can be remotely detected  using spectroscopy, but currently very limited spectral data is available for these molecules.

A more in-depth understanding of planetary environments  through the interpretation of both archival and future observational data, will require spectral data on all relevant atmospheric molecules. To follow-up potential phosphine detections in Venus and exoplanets will similarly require in-depth analyses of the wider context of these atmospheres, which in turn relies on our ability to detect the \pmol{} that participate in the chemical networks where phosphine is present. Thus, discussions and explorations in this paper pioneer key processes and considerations by which an initial biosignature detection can be followed up, and as a by-product identify a wide variety of opportunities and challenges in the field of spectral detection of unknown chemistry (whether geochemical, photochemical or biochemical) that will be crucial for upcoming explorations of exoplanetary atmospheres.

\pmol{} are particularly interesting in astrobiology due to the phosphorus's ability to create complex organic molecules with unique functionality. Phosphorus plays a universally vital role in cellular metabolism (ATP), storage of genetic information (RNA/DNA), formation of cell membranes (Phosolipids) and in cell regulation (phosphate buffer). Phosphorus, in the form of phosphates (\ce{PO4^{3-}}), plays an essential role in carbon chemistry as it: 1) maintains constant negative charge in biochemical conditions; 2) most phosphates (especially polyphosphates) are thermodynamically unstable and have multiple energetic intermediates that can enable polyphosphates like ATP to act as rechargeable batteries in nearly all cellular metabolism; and 3) it works as an efficient pH buffer with free phosphate in cellular plasma regulating acidity \citep{14Pa.M}. Phosphorus is present in all life on Earth \citep{16Co.P} and is expected to be central to life elsewhere. Therefore, understanding the abundance and chemical form of phosphorus on other planets will play an important role in the search for life beyond Earth \citep{03El.J,20Hi.N}.

Understanding the spectroscopy of \pmol{} has value beyond the search for biosignatures in planetary atmospheres. Although phosphorus is a relatively scarce element, \pmol{} are ubiquitously found throughout the galaxy: various \pmol{}  participate in the convective and chemical cycles of gas giants and are expected to behave similarly in cool stars (e.g., \cite{tokunaga1980phosphine, visscher2006atmospheric, larson1977phosphine, weisstein1996submillimeter}); phosphine and five other \pmol{}  have been detected in circumstellar regions (\cite{agundez2014confirmation,turner1987detection, ziurys1987detection, agundez2007discovery, guelin1990free, tenenbaum2007identification, halfen2008detection}); and also comets with a significant phosphorus content are expected to have \pmol{} in their coma (\cite{agundez2014molecular, crovisier2004composition, macia2005role, kissel2004cometary}.

Despite its relevance for both astronomy and Earth sciences, the rich chemistry of atmospheric phosphorous species is understudied, partially because of the paucity of spectral data for these molecules. Lack of suitable spectral data seriously hinders atmospheric characterisations. Obviously, if there is no spectral data on a molecule suitable for use in astronomy, the spectroscopic detection of the molecule can never be confirmed.  Incorrect assignments are also a strong possibility, particularly with the limited low resolution data that is commonly available from exoplanet observations; it is all too easy to incorrectly assign a spectral signature when the reference data is not sufficiently comprehensive. As an example, the recent confirmation of water on the Moon \citep{20HoLuLiShOr,20ScWi} relied on a less ambiguous spectral signature at 6 $\mu$m -- the H-O-H bend region -- rather than earlier detections at 3 $\mu$m -- the O-H stretch region -- which could have easily been any molecule with an alcohol (-OH) functional group such as methanol. 

Production of high-quality rovibrational or rovibronic spectral data is extremely time-consuming even for small molecules, usually requiring a combined theoretical-experimental approach to achieve accurate frequency and intensity predictions that are reliable across a large spectral region. For a thorough description of the current theoretical approaches including strengths and limitations, see \cite{tennyson2016ab} for diatomic rovibronic spectroscopy and \cite{tennyson2016perspective} for polyatomics. Experimentally, typical laboratory challenges are exacerbated for \pmol{}  due to safety concerns around many phosphorous-containing molecules and the difficulty in obtaining or synthesising pure samples.  This research effort is only reasonable for a relatively small number of molecules, targeted for their importance in known or proposed chemical processes.

An alternative is to use high-throughput methods to produce spectral data for hundreds to thousands of molecules. These procedures can be used to identify groups of molecules difficult to distinguish, screen molecules with strong transitions and provide a database for alternative potential assignments of observed spectral features. High-throughput methodologies often must compromise accuracy for coverage. Nonetheless, they allow for a statistical and pattern-focused analysis of atmospheric and molecular spectra that is mostly out of reach to traditional spectral databases. \cite{19SoPeSe.P} pioneered this data generation by  producing approximate spectral data for more than 16,000 molecules using a functional-group driven approach called {\package RASCALL} that relies primarily on organic chemistry. In this paper, we develop a complementary approach, called CQC, using automated approaches with standard computational quantum chemistry (hence CQC) methods to produce spectral data for more than 900 \pmol{} over a wider spectral range than \package{RASCALL} data.
 
This paper focuses on infrared spectroscopy of gas-phase \pmol{}  and is organised as follows. Section \ref{sec:mol} presents an extensive literature synthesis of potentially volatile \pmol{} that could be spectroscopically observed in planetary atmospheres. Section \ref{IR_spectroscopy} collates and discusses the key existing sources of infrared spectral data for P-molecules along with presenting and analysing our new results for 958 molecules obtained with computational quantum chemistry (CQC). Section \ref{discussions} considers the diverse uses of our new large spectral CQC dataset, discusses the interplay of spectroscopic detections with reaction network and kinetic modelling, and reflects on the interdisciplinary approach adopted in this paper. Finally, in section \ref{conclusions} we conclude with a summary of the key contributions of this paper. 

The scope of this paper is deliberately broad. We aim to identify critical research sub-projects for future detailed analysis, whilst simultaneously (and as importantly) identifying sub-projects with less impact potential. To achieve this broad perspective, we directed interdisciplinary expertise to the specific problem of P-molecule atmospheric speciation and spectroscopy. Significant insights to the problem were contributed from computational quantum chemistry, astronomy, atmospheric chemistry, kinetics and reaction network modelling, experimental spectroscopy, machine learning, geology and origin of life research disciplines. 

\section{Potentially volatile phosphorus-bearing  molecules}\label{sec:mol}

In this section, we tackle the challenging problem of enumerating and prioritising the phosphorus-bearing molecules (\pmol{}) that could be spectroscopically observed in planetary atmospheres. There are two main approaches to this problem:
\begin{itemize}
    \item a targeted approach, developed in sub-section \ref{subsec:reactiontargeted}, that iteratively builds up a list of molecules based on known or proposed chemistry in planetary atmospheres including Jupiter, Earth and Venus; 
    \item a reaction-agnostic approach, developed in sub-section \ref{subsec:reactionagnostic}, that simply enumerates all molecules that fulfil certain criteria.
\end{itemize}

\subsection{\label{subsec:reactiontargeted}Targeted Approach}
Our goal in this section is to identify target \pmol{}  that may be detectable in planetary atmospheres, including species that are predicted to be important for understanding the phosphorus chemistry on Venus. Table \ref{tab:specificinterest} details the small number of atmospheric \pmol{}  that have been explicitly considered.

\begin{table}[htbp!]
\small
    \centering
    \caption{  \label{tab:specificinterest} Non-diatomic potentially gaseous phosphorus-bearing molecules identified in the literature (ref column) as relevant to terrestrial atmospheres. Simplified Molecular Input Line Entry System (SMILES) notation are provided for molecules with six or fewer non-hydrogen atoms. The abbreviations are as follows: \textit{R} means in list of AllMol dataset with \package{RASCALL} data available (excluded molecules are generally intermediates, radicals or contain more than 6 non-hydrogen atoms), \textit{LL} means high resolution line list available, \textit{ExpDB} means present in experimental database other than NIST while \textit{ExpLit} means we have identified spectra in the experimental literature (a non-comprehensive search), with \textit{s}, \textit{l}, \textit{aq}, \textit{g} indicating molecular state of spectra, with \textit{matrix} indicating argon matrix spectra (similar to gas phase).  }
\begin{tabular}{@{}lp{3cm}p{5.5cm}p{2cm}p{4cm}HHHH@{}}
\toprule
Molecule & \package{SMILES}     & Name  & Ref & Spectral Data \\ 
\midrule

\mc{3}{l}{\textbf{Produced by life}} \\
\ce{H3PO4}    & O=P(O)(O)O & phosphoric acid    &  e.g. [1, 6] & R, ExpLit (\textit{aq} [11], \textit{l} [12]) \\
\ce{PH3}& P  & phosphine  & e.g. [1]  & R, LL, ExpDB  \\
\ce{CH5O3P}   & O=P(O)(C)O & methylphosphonic acid  & [6] & R, NIST (\textit{s})  \\
\ce{C2H8NO2P} & O=P(CCN)O   & (2-aminoethyl)phosphinic acid  & [6] & R, ExpLit (\textit{s} [13])  \\
\ce{C2H7O3P}  & O=P(OC)OC  & dimethyl phosphonate  & [6] & R, NIST (\textit{g}) \\
\ce{CH5O4P}   & O=P(O)(CO)O& (hydroxymethyl)phosphonic acid & [6] & R \\
\ce{CH5O4P}   & O=P(O)(O)OC& methyl dihydrogen phosphate  & [6] & R \\

\vspace{-0.5em} \\
\mc{3}{l}{\textbf{Other Phosphorus Oxoacids}} \\
\ce{H3PO3}    & O=P(O)O    & phosphorus acid  & [3]  & R, ExpLit (\textit{l} [12])  \\
\ce{H2PO3}  & OP(O){[}O{]}   & (dihydroxyphosphino)oxidanyl & [6] & None \\
\ce{H2PO3^-}  & OP(O){[}O-{]}  & dihydrogenphosphite& [4] & None \\
\ce{H4P2O7}  &     & pyrophosphoric acid& [1] & None \\
\ce{H5P3O10}  &   & triphosphoric acid & [1] & None \\

\vspace{-0.5em} \\
\mc{3}{l}{\textbf{Phosphorus oxides}} \\

\ce{P4O6}     &      & tetraphosphorus hexaoxide& [1, 3, 9]  & ExpLit (\textit{matrix} [14]) \\
\ce{P4O10}    & & phosphoric anhydride   & [1, 3, 9] & ExpLit (\textit{g} [15]) \\
\ce{P2O5}     & & phosphorus(V) oxide&  [7] & None \\
\ce{PO2} & [O-]P=O & hypophosphite & [1] & None \\

\vspace{-0.5em} \\
\mc{3}{l}{\textbf{Organophosphorus compounds}} \\
\ce{C4H9P}    & C1CCCP1    & phospholane& [2,4] & R \\
\ce{CH5P}     & CP & methylphosphine    & [4] & R, ExpLit (\textit{g} [16]) \\
\ce{C3H7P}    & C/C=C/P    & {[}(E)-prop-1-enyl{]}phosphane & [5] & None \\
\ce{C3H7P}    & C/C=C\textbackslash{}P & {[}(Z)-prop-1-enyl{]}phosphane & [5] & None \\
\ce{C2H5P}    & PC=C & vinyl phosphine    & [5] & R, ExpLit (\textit{g} [17])\\ 
\ce{C3H9P}    & CP(C)C     & trimethylphosphine & [4] & R, ExpDB, ExpLit (\textit{g} [18]) \\
\ce{C2H5P}    & P1CC1 & phosphirane & [5] & R & ExpLit (\textit{g}) \\
\vspace{-0.5em} \\

\mc{5}{l}{\textbf{Intermediates not otherwise specified}} \\

\ce{PH2}& {[}PH2{]}  & phosphino  & [8] & ExpDB (\textit{mw only}) \\

 \bottomrule
\end{tabular}
\\ 
References to relevance of the molecule in table are [1] \cite{20BaPeSe.P}, [2] \cite{19SoPeSe.P}, [3] \cite{20GrRiBa.P}, [4] \cite{19BaPeSe.P}, [5] \cite{95GuJaLa.P}, [6] \cite{16SeBaPe.P}, [7] \cite{06VlKr.P}, [8] \cite{20MoLiWa.P}, [9] \cite{89Kr.P}, [10]. Experimental data references are [11] \cite{10Ru.P}, [12] \cite{20FaFeKr.P}, [13] \cite{72MeWa.P}, [14] \cite{89MiAn.P}, [15] \cite{92KoCoBo.P}, [16] \cite{65Mo.P, 59LiNi.P}, [17] \cite{06BeBePo.P}, [18] \cite{60Ha.P}

\end{table}


Let us first clarify important terminology and phosphorus chemistry concepts. \ce{PO4^{3-}} is the most oxidised form of phosphorus, with a phosphorus oxidation state of $+5$, and is generally present in the atmosphere as phosphoric acid, \ce{H3PO4}. Other forms of phosphorus are considered to be reduced phosphorus, with \ce{PH3} being the most reduced form of phosphorus (oxidation state of $-3$), but not thermodynamically favoured at temperatures below 800 K with low hydrogen-pressure \citep{visscher2006atmospheric} or in oxidising environments like modern Earth where it reacts rapidly with \ce{OH^{.}} and \ce{O^{.}} radicals. Therefore, the dominant forms of phosphorous on a planet will depend on the planet's (bio)geochemical cycles, as well as whether the atmosphere contains reduced (e.g. \ce{H2}, \ce{CO}) or oxidised gases (e.g. \ce{O2}, \ce{H2O}, \ce{CO2}).

Phosphorus compounds are generally categorised as organic (containing carbon) or inorganic (do not contain carbon). Only some of the \pmol{} in the atmosphere are volatile, for example large quantities of inorganic phosphorous are dispersed into the atmosphere as coarse solid particles (aerosols) from dusts or combustion sources \citep{08MaJiBa.P}.  Inorganic (poly)phosphates and several organic atmospheric phosphorous compounds are soluble in water and thus bioavailable.   Additionally, plant activity can emit complex biogenic \pmol{} that aggregate as coarse aerosols, which are insoluble and only transport and deposit organic phosphorous locally, rather than globally \citep{14TiBeBo.P}. 


\subsubsection{\pmol{}  in hydrogen-rich reducing gas giants, e.g. Jupiter, Saturn}

In the reducing environments of Jupiter and Saturn, the most abundant P-molecule is phosphine (\ce{PH3}). Though phosphine is not the most thermodynamically favourable form of phosphorus at temperatures of the atmosphere, phosphine formed in the hot deep layers is brought to the top of the atmosphere through convection. In modelling this phenomena and seeking to understand the phosphorus chemistry of gas giants, \cite{78BaLe.P,94FeLo.P,95BoDoKh.P} considered the abundances of \ce{PH3}, \ce{PH2}, \ce{PH}, \ce{P4O6}, \ce{P4O7}, \ce{P4O8}, \ce{P4O9}, \ce{P4O10},   \ce{PS}, \ce{P2}, \ce{P}, \ce{PO}, \ce{PO2}, \ce{PF}, \ce{PC}, \ce{PCl}, \ce{PN}, \ce{P4} and \ce{P3}; with many of these compounds having very low modelled abundances.  \ce{P4O6} and \ce{P4O10} are particularly notable as they arise often in the literature considering \pmol{} and gas-phase chemistry due to their stability, despite their large molecular weight. \ce{P4O6} has a boiling point of 173.1 $^{\circ}$C, while \ce{P4O10} sublimes at 360 $^{\circ}$C. These properties implies the vapour pressure and thus gaseous abundance of both compounds may be appreciable, especially in higher temperature environments.  It has also been hypothesised that alkyl phosphines, i.e. \ce{PR_1R_2R_3}, may be formed in  hydrogen-rich environment from the photolysis of \ce{PH3} in the presence of hydrocarbons \citep{95GuJaLa.P,97GuLeLa.P}.  










\subsubsection{\pmol{}  expected on Earth}
The speciation of phosphorus in the Earth's atmosphere is quite different than gas giants. Earth's atmosphere is an oxidising environment and therefore the reduced species \ce{PH3} is associated solely with biological and industrial activity \citep{sousa2019phosphine}. Instead, phosphates (\ce{PO4^{3-}}) are most common, with \ce{H3PO4} assumed as the dominant species and the only P-molecule with gas-phase kinetic data \citep{20BaPeSe.P}. In the context of this paper, the most notable thing about phosphorus on Earth is the almost complete absence of gas phase \pmol{}. Most descriptions of Earth's phosphorus cycle (e.g. \cite{20ScBe.P}) completely ignore any atmospheric involvement of \pmol{}, and focus instead on the much more numerous and biologically critical processes by which phosphorus moves through the lithosphere, hydrosphere and biosphere.

The atmospheric impact of \pmol{}  can usually be neglected because most \pmol{}  either have low volatility (such as \ce{P4O10}) and quickly ``rain out" into the hydrosphere, or are highly reactive and are destroyed in Earth's oxidising atmosphere. Consequently, few \pmol{}  are the subject of study in the Earth's atmosphere, and no \pmol{}  are included explicitly in the two most chemically comprehensive Earth atmospheric models, the Master Chemical Mechanism \citep{MCM} and GEOS-chem \citep{GEOS-Chem}.

For the phosphorus that does cycle into the Earth's atmosphere, the conditions are so oxidising that atmospheric budgets have total atmospheric phosphorus (primarily from dust or biogenic aerosols) as generally being oxidised to \ce{PO4^{3-}} and deposited into the Earth's oceans \citep{08MaJiBa.P}. Atmospheric deposition and biological fixation of phosphorus is typically considered negligible in comparison to total phosphorus, and subsequently ignored in terrestrial phosphorus budgets \citep{19ZhLiLi.P}.

Experimentally, the speciation of atmospheric phosphorus on Earth is still poorly understood; typical analytical techniques destroy speciation information \citep{03MoGlEd.P}, such as acidification of samples to pH 1 in spectrophotometry \citep{08MaJiBa.P}. Contemporary techniques are able to distinguish between soluble/insoluble phosphorous and inorganic/organic phosphorous \citep{18ViBoAu.P}, but an exact chemical inventory of these species has not been made. Recently, there has been recognition of plant emissions such as phosphate esters, i.e. \ce{P(OR_1)(OR_2)(OR_3)}, in contributing to atmospheric volatile organic phosphorus, not just to coarse biogenic aerosols \citep{20LiLiTa.P}; but the overall impact of atmospheric organic phosphorous is not widely recognised yet.


A perhaps surprising source of information on potential gaseous \pmol{}  in atmospheres comes from the origin of life literature. Phosphorus is considered an essential component of life, yet dominant phosphorus sources (notably apatite) are only slightly soluble, raising the question of how phosphorus was introduced into the hydrosphere in sufficient quantities to enable life to emerge on Earth (e.g. \cite{91YaWaSa.P,06Schwartz.P}). Studies into the solution to this phosphorus problem - usually volcanoes, lightning, and meteorites -  lead to consideration of some gas-phase \pmol{}. For example, \cite{91YaWaSa.P} discussed the volatilisation of \ce{P4O10} from high temperature apatite in volcanoes; \ce{P4O10} can then be hydrolysed to form phosphates such as \ce{H3PO4}. \cite{06Schwartz.P} considers the production of phosphite \ce{PO3^{3-}} and phosphorus acid \ce{H3PO3} by lightning in volcanoes. \cite{20RiMoSu.P} also proposed that water can react with meteorite mineral to produce organophosphates by reacting with the phosphide species (containing \ce{P^{3-}}) from meteorite mineral enstatite chondrites to produce \pmol{}  with various oxidation states that are then fully oxidised through photochemical reactions to the bioavailable \ce{PO_4^{3-}} form. Ablation of cosmic particles can produce phosphorus gases such as \ce{PO2} that then dissociates to \ce{PO} \citep{20SaDiBo.P}.

\subsubsection{\pmol{}  expected on Venus}
In the observable upper and middle atmosphere, Venus is an oxidising environment due to the high concentration of sulfuric acid - a strong oxidising agent - and the high production rate of oxidising radicals through photolysis \citep{20BiZh.P}. Therefore, \ce{H3PO4} is predicted to be the most dominant P-containing species in the upper atmosphere \citep{03GlEdKu.P}, with some phosphate in the form of dehydration products, e.g \ce{H4P2O7}, \ce{H5P3O10} \citep{20BaPeSe.P}. \ce{H3PO3} concentration is predicted to be negligible, at tens of milligrams across the whole atmosphere  \citep{20BaPeSe.P}. 
 In the lower Venusian atmosphere, \ce{P4O6} is thermodynamically favoured and dominates the chemistry of \pmol{} at this altitude \citep{89Kr.P}, while \ce{P4O10} is disfavoured. 

Overall, similar geological processes are likely to occur in Venus as on Earth as the bulk composition for both planets is expected to be similar \citep{09Treiman.P,13Shellnutt.P}.
Some differences are expected due to the differing atmospheric composition (far less \ce{O2}, more \ce{CO2} and more sulfuric acid) \citep{johnson2019venus}, lack of plate tectonics on Venus \citep{nimmo1998volcanism}, the higher ground temperature \citep{taylor2018venus} and lack of water oceans \citep{taylor2018venus} on Venus. 
The effect of these differences on the atmospheric speciation of P-molecules is unexplored.


\subsubsection{\pmol{} from life on Earth}
Life can produce a much richer range of molecular species than geological processes and has the potential to influence the sources and sinks of molecules to drastically impact the atmospheric composition, e.g. enabling 21\% oxygen on Earth. P-molecule species produced by life \citep{16SeBaPe.P} include \ce{CH5O3P}, \ce{C2H8NO2P}, two structural isomers of \ce{CH5O4P}, \ce{H3PO4} and, of most recent interest, phosphine (\ce{PH3}).

As reviewed by \cite{sousa2019phosphine}, on Earth, atmospheric \ce{PH3} is associated exclusively with life, either through anthropogenic sources (e.g., agriculture), or through its production in anaerobic ecosystems (e.g., lake sediments, marshlands), but has very low abundance (ppt/ppq locally at sites of anaerobic activity). The largest sink for \ce{PH3} on the Earth is destruction by \ce{OH^{.}} \citep{glindemann2005phosphine}, causing a very short \ce{PH3} lifetime measured in hours \citep{sousa2019phosphine}. 

Despite phosphine being present in a range of environments -- almost exclusively anoxic -- the exact basis and mechanism for phosphine formation in nature is not well understood. Early work has reported the production of phosphine from mixed bacterial cultures (mixed acid and butyric acid bacteria) in the laboratory  \citep{00JeMoCr.P}. Pasek and colleagues  \citep{14PaSaAt.P} proposed that  phosphite \ce{[H2PO3]^{-}} \footnote{Note there are differing naming conventions with \ce{HPO3^{2-}} also often known as phosphite.} and hypophosphite \ce{[H2PO2]^{-}} are first produced through microbial metabolism, and these compounds are then converted to phosphine by other mechanisms. \cite{19BaPeSe.P} suggest that, in some environments, it is a combination of phosphate-reducing bacteria and the coupling with phosphite metabolism that results in phosphine release. Several very recent studies are beginning to provide more informed insights into the potential roles of specific microorganisms and pathways in phosphine production. For example, \cite{20FaLvNi.P} indicated that the production of acetic acid via the tricarboxylic acid cycle promoted the production of phosphine. Most recently it was found that the phosphine production was enhanced when the hydrogen levels were increased \citep{21FaNiZh.P}. The authors suggested that phosphine production was promoted with hydrogen as an electron donor (i.e. \ce{H2PO4-} + \ce{H+} + \ce{4H2} $\rightarrow$ \ce{PH3} + 4\ce{H2O}), and it was concluded that both reducing power and excess electrons are necessary prerequisites for the production of phosphine. The activity of the enzyme dehydrogenase was shown to be positively correlated with phosphine production  \citep{21FaNiZh.P}, suggesting that this enzyme’s function in producing electrons and reducing agents contributes to phosphine generation. Furthermore, co-factors such as NADH and riboflavin vitamins were suggested to be key in phosphine production  \citep{19BaPeSe.P,21FaNiZh.P}. Given the limited studies and the debate surrounding the exact pathways and the diverse microorganisms potentially involved in phosphine production, significantly more work is needed in this area on the biological basis for phosphine production.

\subsubsection{P-molecules potentially involved in phosphine production on Venus}

Recently, phosphine has risen to prominence due to its potential as a strong biosignature (with few non-biological sources on temperate planets) and tentative detection on Venus. We refer to the very detailed previous publications for known and proposed geochemical and photochemical networks of phosphine production in exoplanets  \citep{sousa2019phosphine} as well as an indepth consideration on Venus \citep{20BaPeSe.P}.

For this paper, we are interested in enumerating the P-molecules identified in these papers as part of the reaction network involving phosphine.  The photochemical network proposed by \citet{20BaPeSe.P} for \ce{PH3} formation involves many other \pmol{}  in a radical reaction network: \ce{H4PO4}, \ce{H2PO3}, \ce{HPO3}, \ce{HPO2}, \ce{HPO}, \ce{PO2}, \ce{PO}, \ce{PH} and \ce{PH2}, several of which will be transient intermediates.

\subsubsection{Discussion of targeted approach} 
The targeted approach followed in this section helped identify molecules of particular interest that are not obvious to non-specialists, such as \ce{P4O10}, as well as identifying challenges to remote detection such as the relative low volatility of many P-molecules. This approach has the ability to synthesise relevant interdisciplinary knowledge across sub-fields and enhance the common understanding of the scope, context and limitations of existing disciplinary expertise. 

Overall, there is a paucity of modelling work on the atmospheric speciation, kinetics and reaction networks of \pmol{}. Though we can identify some species likely to be important, this poor understanding means that it is desirable to consider a much broader range of potential volatile \pmol{}, as described in the next section, in order to detect the \pmol{}  present in a given atmosphere and facilitate the elucidation of atmospheric phosphorus reaction networks. This broad perspective will be particularly critical for characterising diverse exoplanet atmospheres.

\subsection{\label{subsec:reactionagnostic}Reaction-agnostic Approach}

The search for extra-terrestrial life currently relies on the detection of biosignature gases, i.e. gases produced by life that accumulate in the atmosphere and could be detected remotely \citep{18ScKiPa.P}. A molecule's suitability as a biosignature is impacted by multiple factors (e.g. the host planet's atmospheric chemistry and geology) and so any selection criteria are necessarily broad. There is an unavoidable Earth-centric bias in the search for life in the universe; nonetheless, it is laudable to attempt to approach the search for biosignatures on exoplanets as agnostically as possible. \alert{With that in mind, \cite{16SeBaPe.P} proposed a list of more than 16,000 molecules (hereafter AllMol list) that may be associated with life and are likely to be volatile in the atmosphere of potentially habitable exoplanets.} 


The AllMol list contains molecules with up to six non-hydrogen atoms that are expected to be volatile and stable at Earth's standard temperature and pressure (STP). Volatile molecules were estimated as those with boiling points below 150$^{\circ}$C, as most molecules with boiling points above this temperature are likely to be nonvolatile. Stability was interpreted as molecules being able to remain as pure entities under STP conditions and  not reacting  readily with water.  The cutoff of molecules with up to six non-hydrogen atoms was chosen as it implies volatility for a substantial fraction of molecules, including several molecules that are currently studied as biosignature gases.

The AllMol list contains 2,130 \pmol{}  made of the elements C, N, O, F, S, Cl, Se, Br and I. However, for our quantum chemistry studies, we have excluded molecules containing the rare elements Se, Br and I, as they posed additional computational difficulties that were not worth addressing given the rarity of these elements. Specifically, an initial guess geometry could not be generated for the Se-containing molecules, while Br- and I-containing molecules had much larger computational cost due to the large number of electrons. \alert{For completeness, we have included the water-reactive \ce{PCl3} and \ce{PF3} molecules in our calculations due to their relevance in organophosphorus chemistry, thus leading to a working list containing containing 962 \pmol{}.}

\begin{figure}[htbp!]
    \centering
    \includegraphics[width=0.49\textwidth]{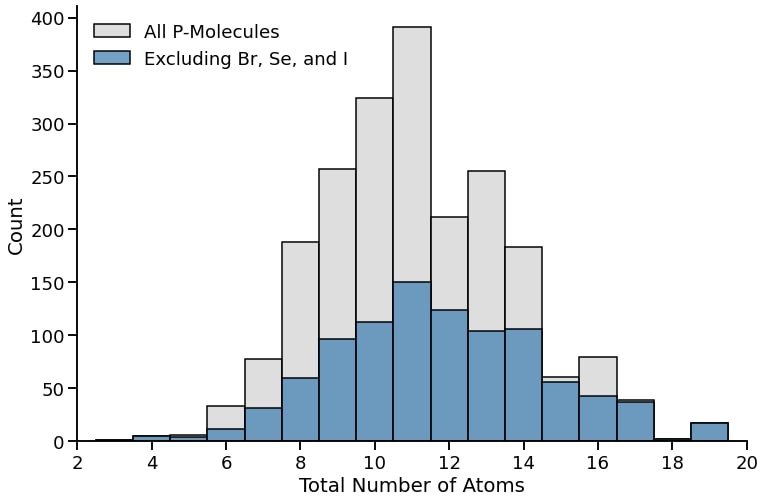}~\includegraphics[width=0.49\textwidth]{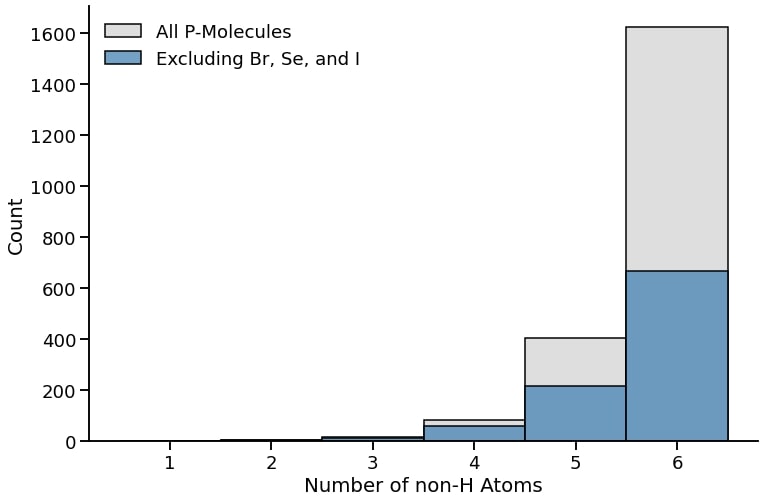}
    \caption{Distribution of AllMol list's 2,130 \pmol{} as a function of total number atoms and non-hydrogen atoms are shown to the left and right top plots, respectively. The total distribution is displayed in grey, with those in blue corresponding to molecules that do not contain Br, Se, and I; our working list of 962 molecules.}
    \label{fig:molecules}
\end{figure}

Figure \ref{fig:molecules} presents the distribution of the molecules present in the full 2,130 \pmol{}  list and our working list with 962 molecules, considering the total number of atoms (left), the total number of non-hydrogen atoms (right). For both lists, most molecules have between 9 to 14 total atoms and five to six non-hydrogen atoms.

\begin{figure}[htbp!]
    \centering
\includegraphics[width = 0.9\textwidth]{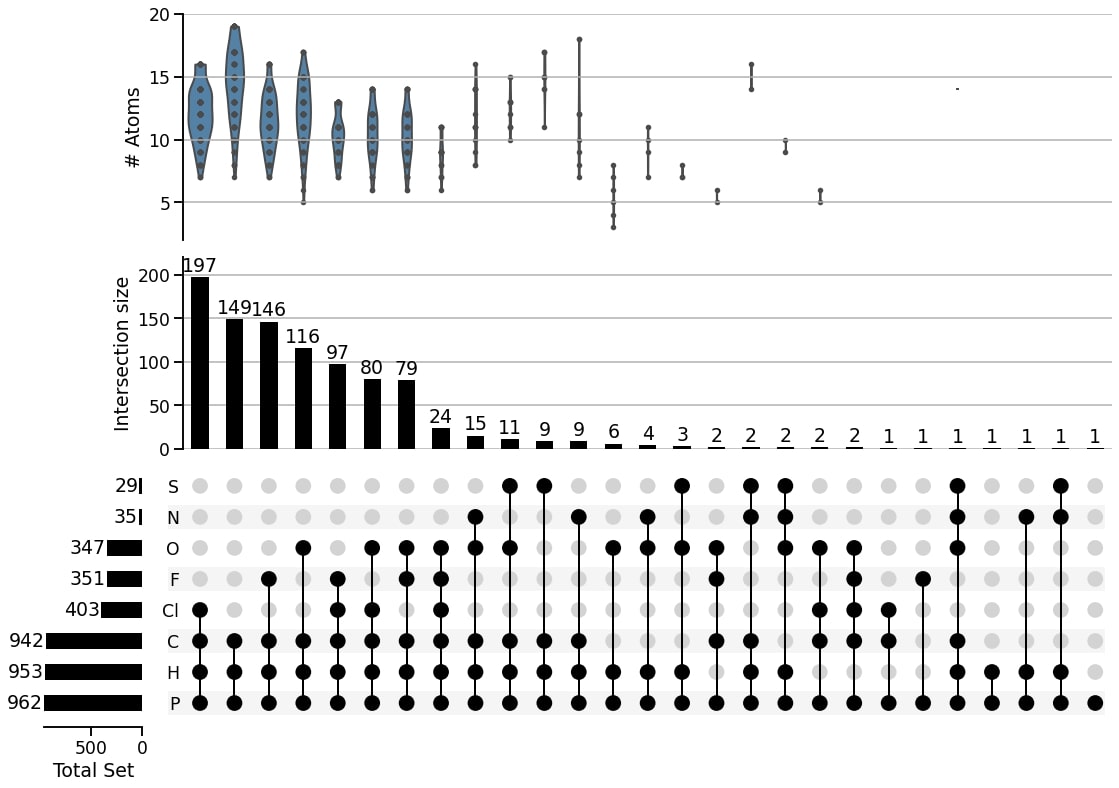}

    \caption{Distribution of elements within our subset of 962 \pmol{}. The horizontal histogram on the left details the number of molecules that contain the corresponding element to the right. The dots create sets of elements that form various molecules, with the counts of those sets shown in the histogram above. For example the first row contains 197 molecules that are made up of only Cl, C, H, and P. The plot above the histogram details the distribution of molecules' size, given by total number of atoms in each set of elements.}  
    \label{fig:atoms}
\end{figure}

For further insight into our working list,  Figure \ref{fig:atoms} shows the count of elemental composition and various combinations of elements, and the number of atoms in each combination of elements. Carbon and Hydrogen are the most abundant elements in \pmol{}  overall, but there are actually more \ce{C_aH_bP_cCl_d} than \ce{C_aH_bP_c} molecules. Almost all our working molecule set contained carbon, i.e. were organic; only 20 molecules were inorganic. 

\begin{table}[h]
    \centering
    
    \caption{Categorisation of \pmol{} within our 962 working set according to the phosphorus bonds in the molecule. Only categories with more than 5 molecules have been included.  }
    \label{tab:class}
    \begin{tabular}{rlHH}
    \toprule
Count & Bond Types \\
    \midrule
    
  \textbf{602}  & \textbf{Organophosphines (Contain P-C and P-H only)} & \\
    236 & 1$\times$P-C, 2$\times$P-H    & 236 (5 diP) \\
252 & 2$\times$P-C, 1$\times$P-H  & 252 \\
107 & 3$\times$P-C  & 107 \\
6 & Contains Aromatic P-bearing ring \\
    \vspace{-0.5em} \\
    
\textbf{203}  &   \textbf{Phosphine oxides (Contain P=O, no P-O)} & \\
61 & 1$\times$P=O, 2$\times$P-C, 1$\times$P-H & 61 \\
58 & 1$\times$P=O, 1$\times$P-C & 58 \\
54 & 1$\times$P=O, 1$\times$P-C, 2$\times$P-H & 52 \\
18 & 1$\times$P=O, 3$\times$P-C & 18 & phosphine oxides \\
        \vspace{-0.5em} \\

\textbf{73} & \textbf{Contains P=O and P-O bond} & 75 \\ 
31 & 1$\times$P=O, 1$\times$P-O, 1$\times$P-C, 1$\times$P-H & 31 \\
13 & 1$\times$P=O, 1$\times$P-O, 2$\times$P-C (phosphinates) & 13 & phosphinates \\
9 & 1$\times$P=O, 2$\times$P-O, 1$\times$P-C (phosphonates) & 9 & phosphonates \\

        \vspace{-0.5em} \\

\textbf{61} & \textbf{Phosphites (Contains P-O, no P=O bond)} \\
24 & 1$\times$P-O, 1$\times$P-C, 1$\times$P-H & 24 \\
8 & 2$\times$P-O, 1$\times$P-C & 8 \\
6 & 1$\times$P-O, 2$\times$P-C & 6 \\

        \vspace{-0.5em} \\

& \textbf{Other Potentially Overlapping Categories} \\
269 & No P-H bonds \\
42 & No P-C bonds \\
26 & Contains P=S and/or P-S bond & 26 \\
21 & Contains P-N bond & 21 \\
7 & Contains Aromatic P-bearing ring & 8 \\

\bottomrule
\end{tabular}

\end{table}

Another useful categorisation is to consider the bonds to the phosphorus atom within our working set of 962 \pmol{}; broadly speaking, this determines the class of the compound. These statistics are detailed in Table \ref{tab:class}. The dominant class were organophosphines (602/962) in which the phosphorus atom was bonded only to carbon or hydrogen.  Phosphorus-oxygen single or double bonds were the other key phosphorus bond types, with 203 molecules with P=O only, 61 with P-O only and 73 with both P=O and P-O bonds. A small number of compounds have P-S, P=S and/or P-N bonds.  Only 42 compounds had no P-C bonds at all. 

Despite the large number of molecules considered, the constraints imposed when constructing AllMol exclude many molecules considered in our targeted approach, especially large phosphate oxides such as \ce{P4O10} and radicals such as the diatomics PN, PC, PO and PH. Therefore, there is surprisingly little overlap between our targeted molecule set and our working list of 962 \pmol{}. This small overlap probably occurs due to the sparsity of gas-phase \pmol{} in Earth-conditions and would not be expected for many other elements. 

\section{Infrared spectroscopy}
\label{IR_spectroscopy}

Infrared (IR) spectroscopy is currently the technique of choice for future searches of extrasolar biosignatures in planetary atmospheres \citep{schwieterman2018exoplanet}. The successful identification of molecules requires available reference spectroscopic data. Therefore, in this section we consider pre-existing experimentally-derived data, RASCALL-generated data based on functional group decomposition and the newly generated CQC spectral data based on computational quantum chemistry (CQC) calculations for the \pmol{} considered.

\subsection{Existing Experimentally-derived Data}

The infrared spectral data available for \pmol{} are relatively sparse, especially in the gas-phase. There are three main sources of data: (1) line lists of spectral positions and intensities, (2) experimental databases usually containing only un-digitised image spectra (often in liquid phase), and (3) individual papers. Of these, only line lists provide astronomers with accessible data in a format suitable for molecule detection. 


Extensive line lists containing spectral line positions and intensities in the infrared spectral regions are available for 11 \pmol{}. These data are generated individually for each molecule by combining the best available experimental data and \textit{ab intio} CQC calculations, with the latter particularly necessary for dipole moments. There are two broad methodological approaches: the variational approach solving the nuclear motion Schr\"{o}dinger equation on an explicit potential energy surface, and the empirical approach where model Hamiltonian constants are used. Specifically, line list data is available in the centralised ExoMol database \citep{exomol2,exomol3,exomol.empirical} in a standardised format for the following \pmol{}: using the ExoMol variational approach \citep{tennyson2016ab,tennyson2016perspective} for \ce{PH3} \citep{exomol.PH3_room,SAlTY.ph3}, \ce{PF3} \citep{PF3.ExoMol}, PN \citep{exomol.PN}, PH \citep{exomol.PH},  \ce{PO} and \ce{PS} (both in \cite{exomol.POPS}), \ce{cis-P2H2} and \ce{trans-P2H2} (both in \cite{exomol.P2H2})  and using the MoLLIST empirical approach \citep{MoLLIST.Bernath} for \ce{CP} \citep{14RaBrWe.CP}. 
Alternative line list data for phosphine are also available in various sources, e.g. HITRAN \citep{hitran2016}, TheoRETs \citep{09NiHoTy.P,14NiReYy.P} and GEISA \citep{GEISA}.

Experimental infrared spectral absorption cross-sections have been collated mainly by national institutes such as the USA National Institute for Standards and Technology (NIST, \citealt{NIST_Infrared}) and the Japanese National Institute of Advanced Industrial Science and Technology (AIST\footnote{SDBSWeb : https://sdbs.db.aist.go.jp (National Institute of Advanced Industrial Science and Technology,date of access)}). NIST's extensive database of spectral data contains cross-sections with a wide range of accuracy, resolution, and instrumental set-ups. For hundreds of molecules, NIST is the only source of spectral data, and as such mistakes can go unnoticed for many years (e.g., liquid state spectra mislabelled as gas phase, incorrectly assigned spectra or vibrational modes \citep{19SoPeSe.P}). AIST's Spectral Database for Organic Compounds has a useful feature to sort by molecular weight as a proxy for volatility as well as a helpful ability to search by spectral features in a given spectral range. 
For the \pmol{} considered here we have identified two (\ce{C2H7O3P} (O=P(OC)OC), \ce{CH5O3P} (O=P(O)(C)O))\footnote{Notation in parenthesis corresponds to the SMILES code for each molecule.} matches in the NIST database, and three (\ce{C2H7O2P}	(O=P(O)(C)C), \ce{CH6NO3P} (O=P(O)(CN)O), \ce{C3H8ClOP} (O=P(CCl)(C)C)) in AIST. 

Often when infrared data cannot be found in databases or linelists, they may still be found in individual papers. However, in the absence of a centralised database, identifying and processing data for individual molecules is time-consuming due to the diverse literature and poor data digitisation. For some target \pmol{}, we performed a non-exhaustive literature search for experimental data. Usually, the  spectral data was contained solely in figures, and digitisation software would be needed. 
Papers containing experimental infrared data from the twentieth century, when a large body of these papers were written, often suffer from problems of saturation (concentration too high for linear absorption), low resolution, and insufficient detail to obtain absorption cross-sections. These problems can result in intensities that are inconsistent with lower pressure measurements, the fine structure being obscured, and difficulty assessing the abundance of the molecule being identified. In the case of more transient molecules generated by pyrolysis, photolysis or \textit{in situ} reaction, the target species may be produced in a mixture of gases, its’ partial pressure unknown and the spectra affected by bands from other species present.  Thus, experimental data can be used for visual identification of molecules but is generally unsuitable for use by astronomers. 

Data useful for astronomical purposes can be obtained by measurements of the gaseous, low-pressure, infrared spectra of molecules at the very low temperatures achievable in a jet expansion (representing the interstellar medium) and room temperature (representing temperate potentially habitable planets). This data should be produced at high resolution in the full mid-infrared range and distributed in digitised format either as a spectra or as individual line identifications. Typically, we would expect the difficulty of the experiment to depend primarily on the ease of acquiring a pure gaseous sample of a molecule. Stable molecules with a high vapour pressure that can be bought from commercial chemical companies would usually be measurable in low-resolution in a day though high resolution scans would take a few weeks, especially if data is taken for multiple temperatures. 
Unstable molecules or those not commercial available would require more extensive synthesis. 

\subsection{Predicted Spectra from Functional Group Decomposition: RASCALL} 
\package{RASCALL} was the first approach to address the large deficiencies in  infrared spectral reference data for atmospheric molecules by mass data production using computational approaches. The \package{RASCALL} program produces spectral data stored in the \package{RASCALL} database; a living document constantly updated \citep{19SoPeSe.P}. Currently, the database contains spectral data for 15,477 out of the 16,368 AllMol molecules, with a total of 201,985 fundamental frequencies. The \package{RASCALL} database contains spectral data for 1,992 P-molecules with 44 different functional groups, only four of which specifically consider P atoms (namely \ce{P-H} bend and stretch, \ce{P=O} stretch, and \ce{(O=)PO-H} stretch).

\begin{figure}[htbp!]
    \centering
    \includegraphics[width = 0.9\textwidth]{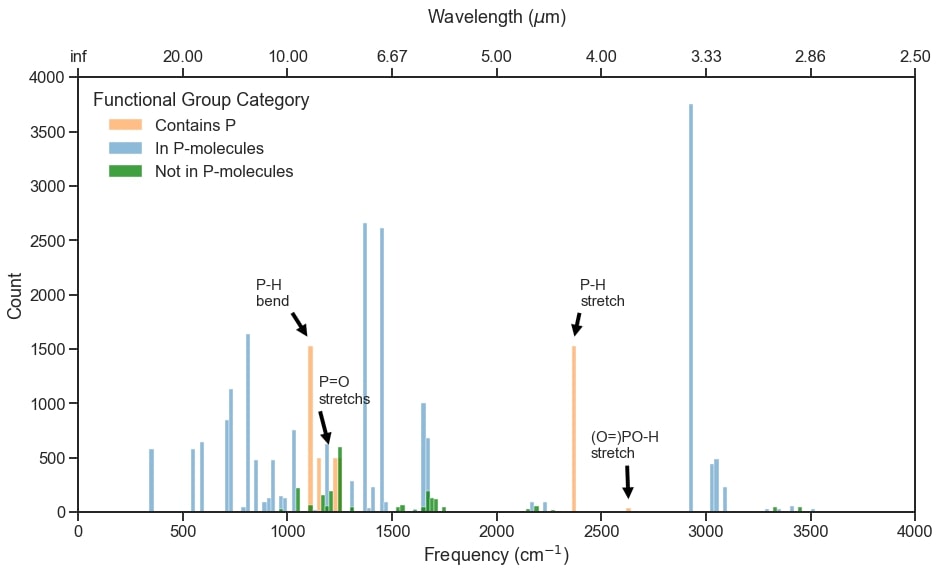}
    \caption{Frequency distribution for all functional groups within \package{RASCALL}. The blue bars correspond to the functional groups present in all 1992 \pmol{}  in \package{RASCALL} that do not involve P and the orange bars are functional groups containing P. The green bars correspond to functional groups that are not included in any of the \pmol{}  present in RASCALL. Data bins of 20 \cm{} has been selected for visibility.}
    \label{fig:RA_fg-p}
\end{figure}

Figure \ref{fig:RA_fg-p} illustrates the functional groups frequency distribution across the \package{RASCALL} data, highlighting those frequencies corresponding to \pmol{}. The P-H stretch and bends are the most ubiquitious P-molecule-specific functional group, with significant numbers of P=O stretches. Of particular interest is the sparsity surrounding the P-H stretch functional group at 2360 \cm{}. This region could represent an interesting signal to look at when searching for molecules with the P-H functional group. The main contaminant will be  \ce{CO2} which has a strong absorption peak at 2,350 \cm that, in high abundances and at low resolution, can obfuscate the spectral feature from the P-H stretch. However, this \ce{CO2} spectral band is usually much more narrow than that caused by P-H stretch, allowing for multiple strong transitions in the wings of the P-H band to be detectable (e.g., as is the case between \ce{PH3} and \ce{CO2}), especially when the P-H stretch frequency is shifted slightly in different environments.  The figure also shows that the majority of functional groups in the data set are shared by molecules with and without phosphorus, with the most prominent one corresponding to the C-H stretch near 3,000 \cm. 

\package{RASCALL} is a computational method that does not utilise quantum chemistry but relies on structural chemistry, especially on functional group theory, to efficiently produce approximate molecular spectral data for arbitrary molecules \citep{19SoPeSe.P}. As functional groups account for characteristic spectral features, \package{RASCALL} estimates the contribution of each functional group present in a given molecule to generate a first approximation to the molecule’s vibrational spectrum. The spectrum given by \package{RASCALL} is composed of the approximate vibrational frequencies of the molecule’s most common functional groups together with the qualitative intensity for each frequency. The functional group database contains more than 100 functional groups and is also a living document, updated as new spectrally active functional groups are identified.


\package{RASCALL} is an extremely quick and powerful approach, but the functional group approach has some inherent limitations. Most notably, the approximate spectra predicted by \package{RASCALL} are based on identified functional groups without taking into account their neighbouring atoms and bonds. This functional group approach makes it nearly impossible to predict the non-localised vibrational modes in the fingerprint spectral region from 500 -- 1450 \cm{}, and restricts the accuracy of local mode predictions in diverse environments. 

Ongoing \package{RASCALL} updates will expand functional group definitions to help address this weakness.  For example, consider the O-H stretch region near 3,600 \cm{} in Figure \ref{fig:RA_fg-p}, where there are few spectral features in our data. As the spectral behaviour of the O-H stretch strongly depend upon the remaining atoms and bonds in the molecule, and consequently vary widely between molecules, \package{RASCALL} does not consider it a single functional group. Instead, \package{RASCALL} uses a categorisation criteria based on different O-H sub-groups that must be considered individually to provide more realistic O-H stretch frequencies. Currently, the \package{RASCALL} database has categorised only a small portion of the O-H variants and those affecting \pmol{} have not been included yet. 

\package{RASCALL} currently only provides qualitative intensities ranging from 1 to 3, representing weak to strong absorption, respectively; this is an area of active method development. 



\subsection{\label{subsec:IRdata}Large-scale computational quantum chemistry data generation: CQC Approach}

\subsubsection{Method}
\label{method_sec}
An alternative to the \package{RASCALL} approach is to use standard computational quantum chemistry (CQC) approaches to directly solve the Schrodinger equation (within a given approximation) and predict vibrational frequencies and intensities of input molecules. Our goal here is to develop the first version of the harmonic CQC-H1 procedure: a high-throughput, largely automated, reliable approach that can be used for hundreds to thousands of molecules by taking as input the molecule's Simplified Molecular Input Line Entry System (SMILES) notation to produce computationally-derived infrared spectra.


Initial molecular geometries for all 962 P-molecules were obtained from SMILES code through a Python script utilising the {\package RDKit}  \citep{RDKit.P}, {\package ChemML}  \citep{20HaViAl.P} and {\package ChemCoord}  \citep{17We.P} libraries. 

Harmonic frequency and intensity calculations for our CQC-H1 approach were performed under the standard double-harmonic approximation utilising the $\omega$B97X-D hybrid functional \citep{08ChHe.P, 16AlFa.P} together with the augmented def2-SVPD basis set \citep{05WeAh.P, 10RaFu.P}. This model chemistry combination (i.e. hybrid functional/double-zeta basis set augmented with diffuse functions) was chosen as it reproduces reliable dipole moments \citep{20ZaMc.P}, a key component for vibrational intensities, and $\omega$B97X-D represents a good general purpose hybrid density functional \citep{goerigk2019trip}. Harmonic frequencies were also scaled using a multiplicative scaling factor of 0.9542 \citep{15KeBrMa.P}. All calculations were performed with the Gaussian 16 quantum chemistry package \citep{g16}.


The initial geometries for all 962 \pmol{} were optimised using a tight convergence criteria (maximum force and maximum displacement smaller than 1.5x10$^{-5}$ Hartree/Bohr and 6.0x10$^{-5}$ {\AA}, respectively) and an ultrafine integration grid (99 radial shells and 590 angular points per shell). For approximately 50 molecules, the jobs did not converge to a minima with the automated approach and needed manual intervention, most commonly recomputing an input geometry in Avogadro \citep{hanwell2012avogadro}. Four molecules were excluded from our analysis due to geometry convergence problems (see sub-section \ref{limitations}), leading to a total of 958 P-molecules considered in our CQC-H1 approach.




\subsubsection{Results}

\begin{figure}[htbp!]
    \centering
    \includegraphics[width = 0.9\textwidth]{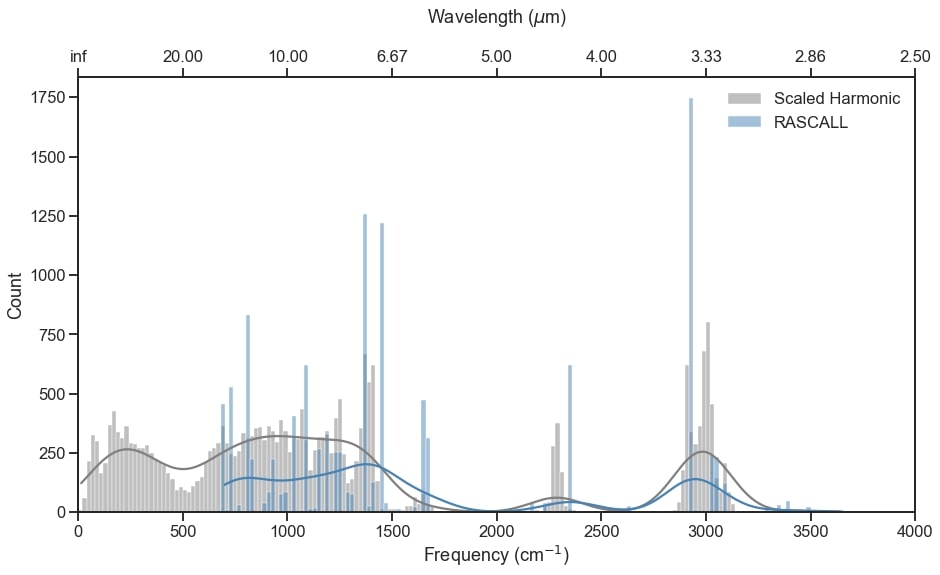}
    \caption{Frequency distribution across the \package{RASCALL} (blue) and the $\omega$B97X-D/def2-SVPD scaled harmonic calculations (grey) data for 868 molecules. A bin width of 20 \cm{} has been selected for visibility. The solid line on top of the histograms represents an estimate of the probability distribution for the data.}
    \label{fig:IR_harm-RA}
\end{figure}

Figure \ref{fig:IR_harm-RA} presents the frequency distribution of the scaled CQC-H1 harmonic frequencies compared to the \package{RASCALL} data, considering the 868 molecules with data available from both sources; incorporating CQC-H1 data from all  958 molecules (total of 28,152 frequencies) produces a similar frequency distribution. 
The fundamentals bands predicted by the harmonic calculations are predominantly found in the region below 500 \cm{}, within 600 -- 1,400 \cm{} (the fingerprint spectral region) and in the 2,900 -- 3,100 \cm{} domain characterised mostly by the C-H stretches. Like the \package{RASCALL} data, our calculated harmonic data also exhibits a small number of signals between 2,000 and 2,700 \cm, apart from the signals around 2,360 \cm, corresponding to the P-H stretch signal. 

Figure \ref{fig:IR_harm-RA} highlights how the CQC-H1 approach, unlike \package{RASCALL}, can differentiate the frequencies at which functional groups absorb based on the specific chemical environment surrounding that functional group. For example, \package{RASCALL} places all C-H stretches at a particular frequency value (prominent blue bar at 2,923 \cm{}), whereas the scaled harmonic calculations for this functional group result in frequencies that are spread over a larger frequency window. The figure also demonstrates the capability of the quantum chemistry calculations to provide data in the fingerprint region of the spectrum (500 -- 1450 \cm{}), as all normal modes are computed (by comparison, RASCALL only predicts around 45\% of the normal modes). Calculation of normal mode frequencies in the fingerprint region poses a fundamental and probably insoluble challenge to the \package{RASCALL} approach, as these fingerprint modes involve motion of large portions of the molecule rather than the movement of isolated functional groups.
\begin{figure}[htbp!]
    \centering
    \includegraphics[width=0.8\textwidth]{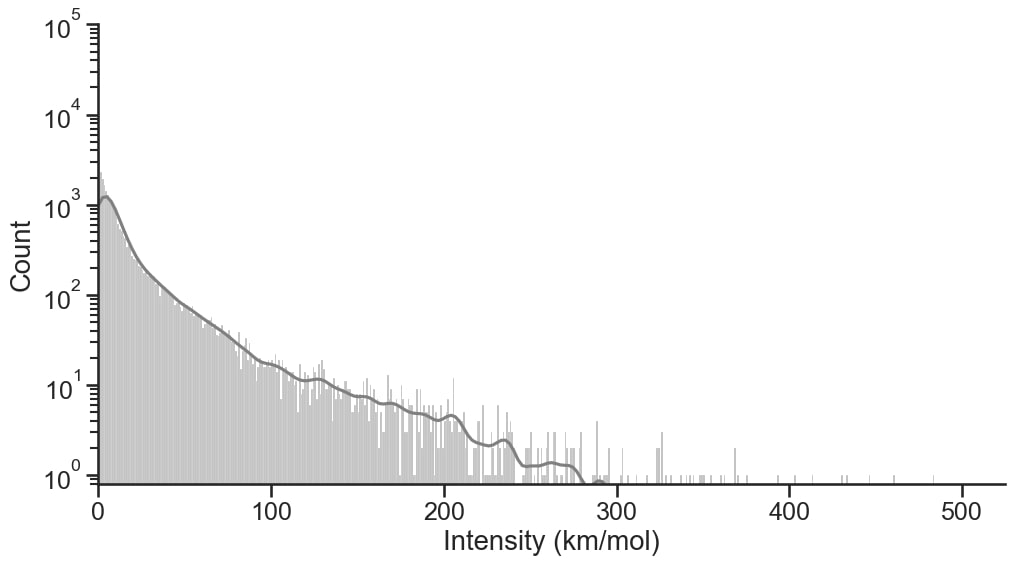}
    \caption{Distribution of intensities for harmonic scaled (958 molecules, 28,152 frequencies). A bin width of 1 km/mol has been selected for visibility. The solid line on top of the histogram represents an estimate of the probability distribution for the data. }
    \label{fig:QM_all}
\end{figure}

Figure \ref{fig:QM_all} shows a logarithmic scale count of the intensity distribution for the harmonic calculations. The count of molecules decreases exponentially with larger intensity values, with a median intensity of 7.5 km/mol.

\begin{figure}[htbp!]
    \centering
    \includegraphics[width = \textwidth]{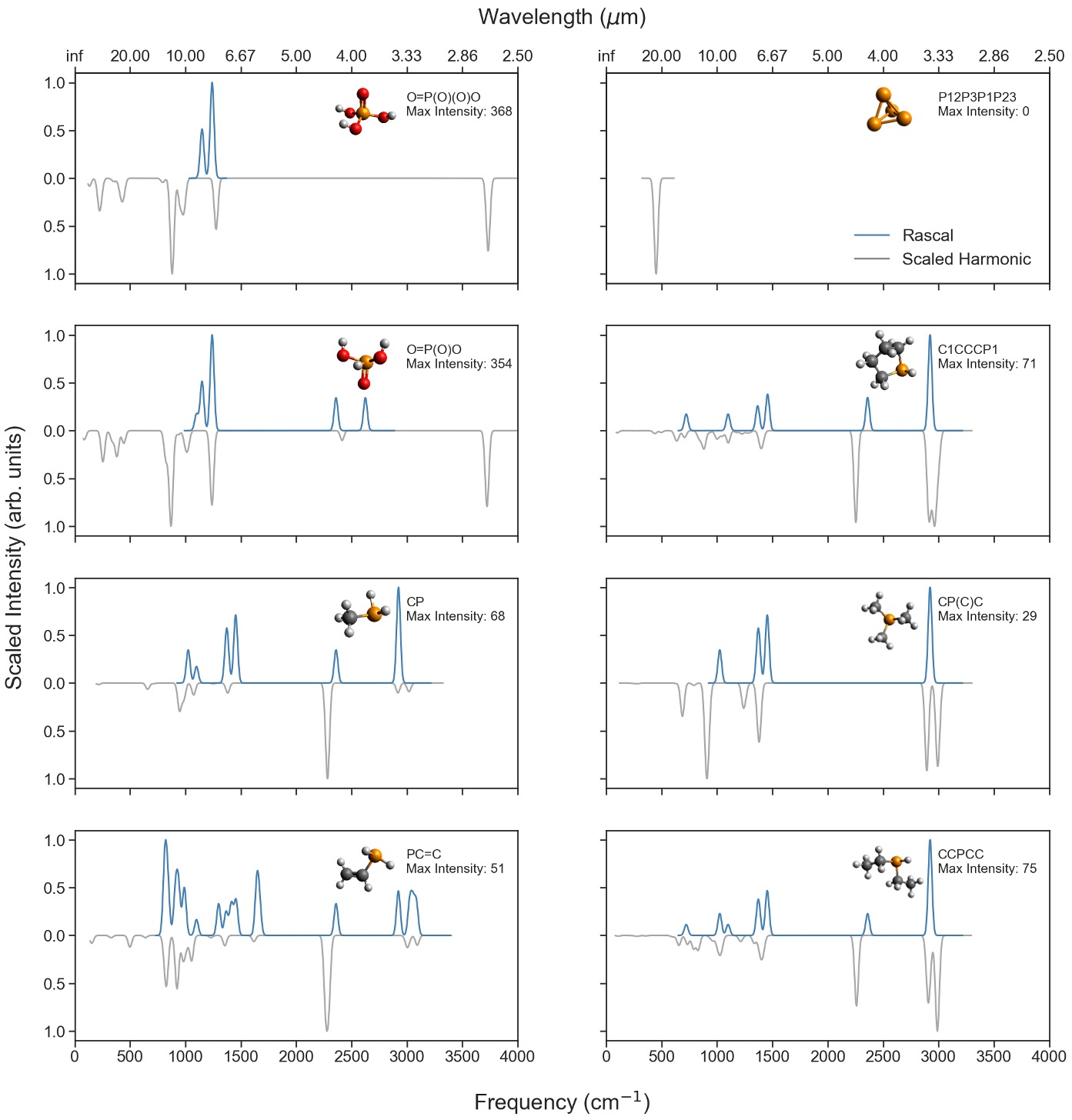}
    \caption{Comparison of the \package{RASCALL} (blue) and scaled harmonic (grey) quantum chemistry data available for eight of the \pmol{}  mentioned in Table \ref{tab:specificinterest}. The \package{SMILES} code is presented for each molecule as well as the largest values for the predicted intensities (km/mol) with the harmonic calculations.}
    \label{fig:interest}
\end{figure}

\begin{figure}[htbp!]
    \centering
    \includegraphics[width = \textwidth]{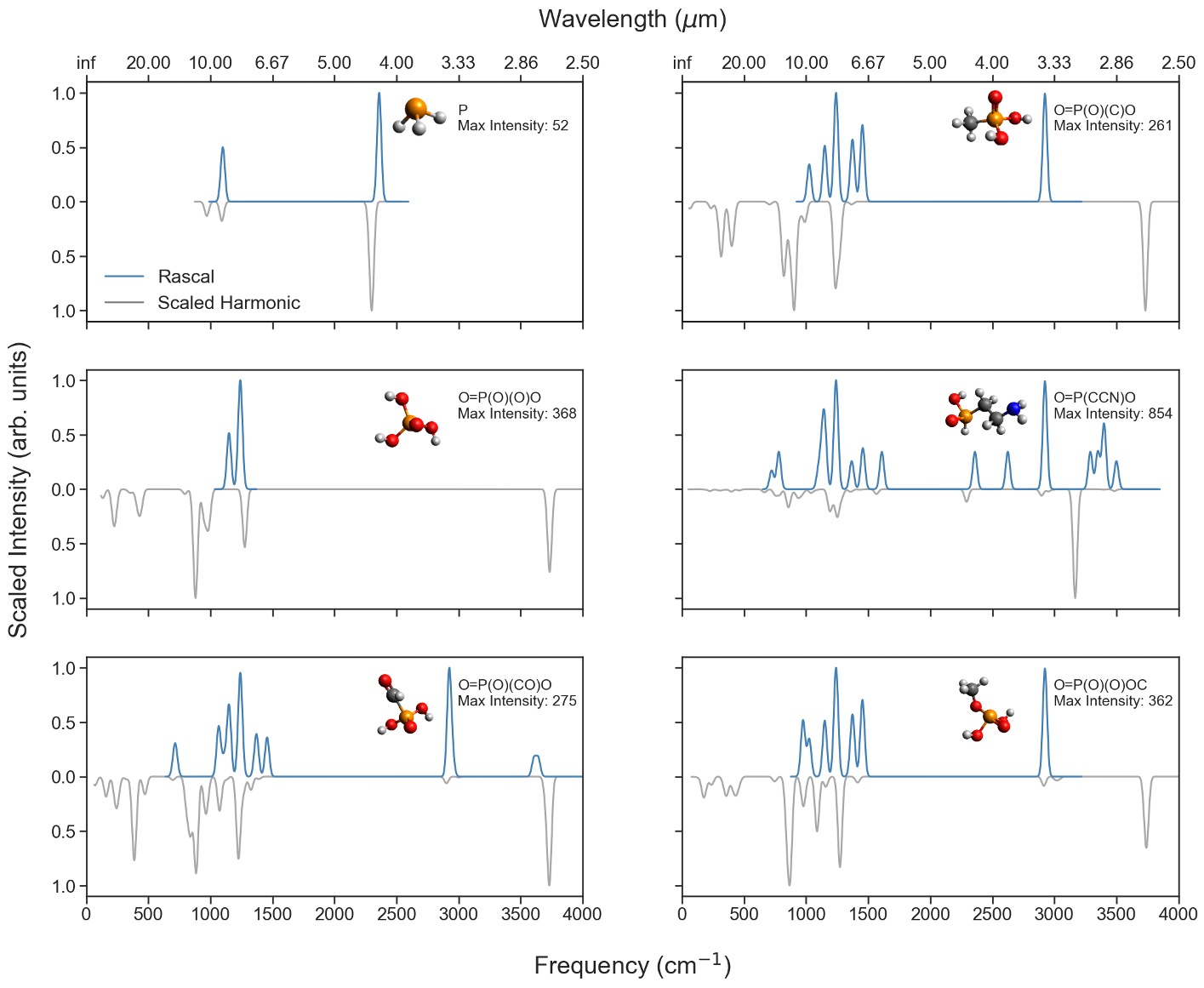}
    \caption{Comparison of the \package{RASCALL} (blue) and scaled harmonic (grey) quantum chemistry data for six \pmol{}  produced by life (according to the AllMol list). The \package{SMILES} code is presented for each molecule as well as the largest values for the predicted intensities (km/mol) with the harmonic calculations.}
    \label{fig:life}
\end{figure}

To further illuminate the differences in spectral predictions,  Figures \ref{fig:interest} compares the \package{RASCALL} and CQC-H1 vibrational spectra for a selection of \pmol{} relevant to planetary bodies, and Figure \ref{fig:life} does the same with \pmol{} formed through biotic processes (see Table \ref{tab:specificinterest}). The SMILES code for each molecule as well as the maximum intensities for the harmonic data is shown in the figure and indicates the vertical scaling of the data. 

Across the molecules presented in these two figures, different degrees of agreement can be observed between \package{RASCALL} and the CQC-H1 data. Overall, for most molecules there is clear semi-quantiative agreement in the location of peaks across both sources of data, while \package{RASCALL} often overestimates the intensity of weak bands, especially around 3,000 \cm. As an example, the \package{RASCALL} spectrum for methylphosphonic acid (O=P(O)(C)O top right of Figure \ref{fig:life}) shows a high intensity peak for the C-H stretch with nothing shown in the harmonic CQC-H1 data as the band intensity is significantly low, highlighting the limitations in \package{RASCALL}'s intensity approximations. Regarding the band positions, several of the subplots in both figures show a shift of more the 20 \cm{} in the \package{RASCALL} data corresponding to the P-H stretch (around 2,360 \cm{}). This likely arises from inadequacies in the P-H frequency data in \package{RASCALL} and could be easily corrected with an update using our new P-molecule CQC-H1 data. These figures also provide further evidence of the current deficiencies in the treatment of O-H stretches in RASCALL.


\begin{figure}[htbp!]
    \centering
    \includegraphics[width = 0.95\textwidth]{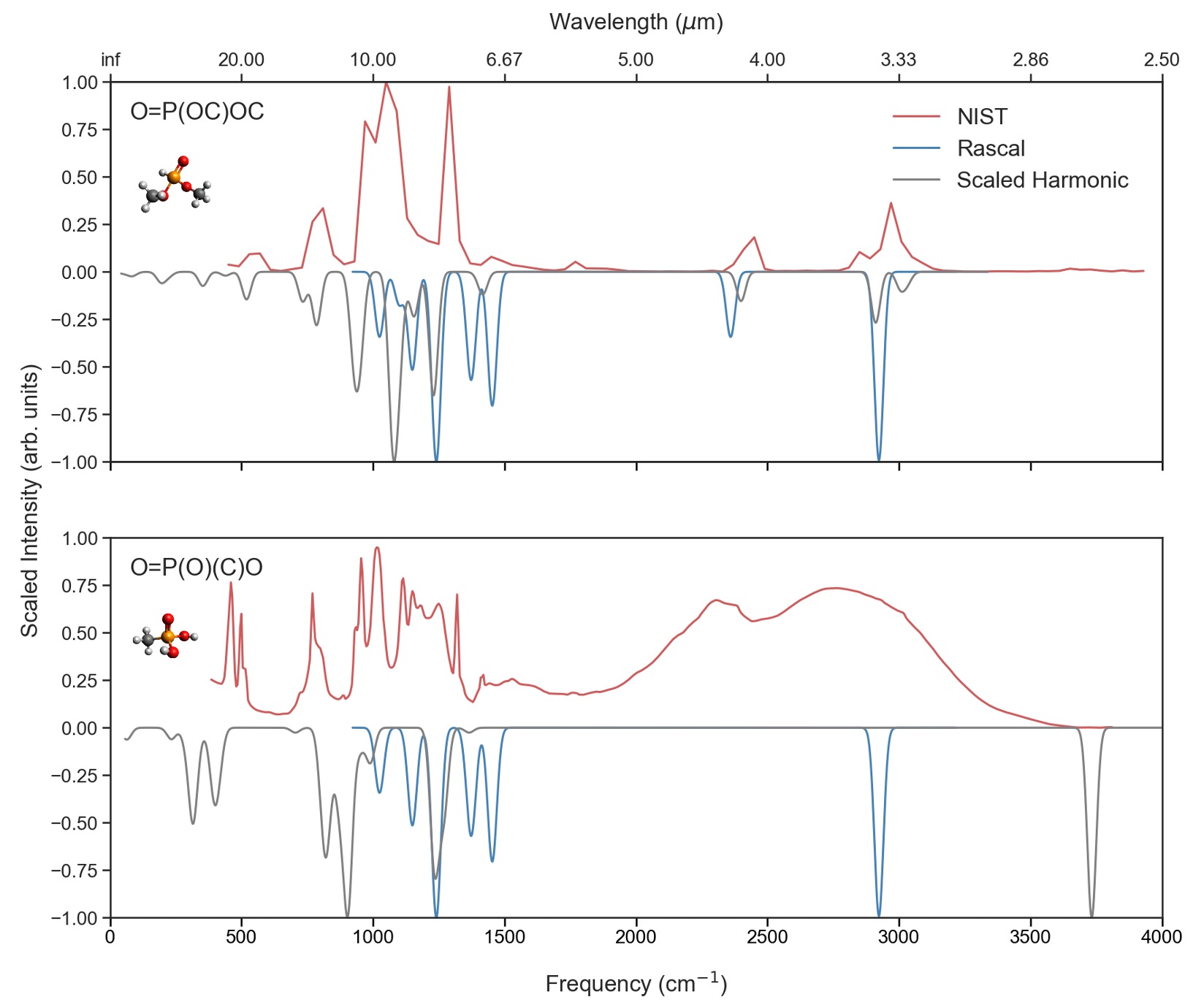}
    \caption{A comparison of the RASCALL, quantum chemistry and experimental data for two \pmol{}  for which experimental data is available. The figure also presents the \package{SMILES} code for each molecule.}
    \label{fig:NIST}
\end{figure}

Figure \ref{fig:NIST} reports the vibrational spectra for two \pmol{}  for which both theoretical and experimental data from NIST is available. The top figure shows an overall fair agreement between the different sources of data, especially in the C-H stretch region where both the NIST and scaled harmonic CQC-H1 data are very alike. In the fingerprint domain (500 -- 1450 \cm{}), the agreement is somewhat less obvious, but still a qualitative similarity is found between the NIST and CQC-H1 data. RASCALL, as previously stated, performs less accurately in this area. On the other hand, there is poorer agreement in the bottom figure as the data collected from NIST corresponds to the solid-phase spectrum for methylphosphonic acid (O=P(O)(C)O) as gas-phase spectrum is not available (or at least not easily accessible). We can see that \package{RASCALL} only provides data for the C-H stretches, disregarding the O-H stretches present in the molecule. Though the scaled harmonic calculations do supply frequencies for both the C-H and O-H stretches, the calculated intensities for the C-H stretches are significantly lower and they are therefore overshadowed by the other frequencies. In the fingerprint region, the agreement between the experimental and scaled harmonic data is somewhat better, with the scaled harmonic frequencies being slightly off and missing some bands. This discrepancy could be due to anomalous frequencies or hydrogen bonding manifesting in the solid-phase spectrum. A gas-phase spectrum would be preferred for a more meaningful comparison. 

Something particularly important to highlight from this analysis is that among all the 958 \pmol{}  considered in our study, we could only find easily accessible reference spectroscopic data for the two molecules illustrated in Figure \ref{fig:NIST}, justifying the necessity of supplemental sources of reference data.

\subsubsection{Consideration of Anharmonic Vibrational Treatment}

Further improvements to our harmonic CQC-H1 approach may be achievable by performing the calculations with the more expensive and complete Generalised Vibrational Second-order Perturbation Theory (GVPT2) approximation \citep{05Ba.P, 19PuBlTa}; an anharmonic method that allows the calculation of overtones and combination bands along with the fundamental frequencies. We tested this approach (hereafter named as CQC-A1) by calculating anharmonic frequencies and intensities for 250 smaller molecules in our dataset, finding significant issues.


Specifically, substantial deviations were found between the scaled harmonic (CQC-H1) and anharmonic (CQC-A1) fundamental frequencies across the molecules tested. These deviations were predominately observed in two cases: (1) low-frequency transitions with small force constants, where differences of up to a factor of 50 between the scaled harmonic and anharmonic frequencies were found; and (2) the P-H stretch frequencies, where the anharmonic frequencies are 200 - 700 \cm{} or higher over the scaled harmonic ones. The first issue is well-known for large amplitude vibrations (LAMs) and is an inherent limitation of perturbative approaches, while the second issue is far more concerning and can be traced back to deficiencies in the density functional to compute higher-order derivatives of the potential energy, as recently noted by \cite{20BaCeFu}.

The  theoretical foundations, technical specifications and analysis of our anharmonic results  are provided in Appendix A, including a discussion of these two key failures and their likely causes.

For the purposes of the main article's objectives, we can conclude that (1) anharmonic GVPT2 calculations are not yet suitable for automated high-throughput calculations due to the prevalence of unexpected anomalous unreliable results and (2) to establish the best anharmonic treatment will require careful testing against experimental frequencies and some criteria that identifies when calculations are unreliable. 


\subsubsection{Challenges, Limitations and Future Directions}
\label{limitations}

The calculation of vibrational spectra for all \pmol{} has various challenges and limitations that are worth discussing.

Our automated approach for obtaining molecular geometries is generally successful, but some issues and limitations were found with its performance. First, as the current version of the libraries used in our automated script for initial geometries does not support the optimisation of molecules containing Se, these molecules were excluded from our final data set. Second, the generated geometries often optimised to saddle-points (indicated by one or more imaginary frequencies) and needed to be manually corrected. As a matter of fact, one of the \ce{C4H7P} isomers (CC\#CPC) had to be removed from our working set due to the prevalence of imaginary frequencies in the calculation despite testing the available options to deal with this problematic; future work will attempt to automate this process to enable high throughput calculations. Finally, for three molecules (namely \ce{C2H4FO2P} (O=P1(O)C(F)C1), \ce{C3H4ClOP} (O=PCC=CCl) and \ce{C2H5O2P} (O=P1(O)CC1)), the geometry optimisation procedure led to their decomposition in the final state, due to their inherent instability. These molecules were identified through their anomalous vibrational partition functions that were calculated for a future application, but could have easily been missed.

Perhaps the most significant limitation of our current method is that only one conformer, as calculated by our automated approach, was considered. However, other conformers may certainly have lower energies than the ones used in our calculations. This limitation will be addressed in the future by consideration of multiple conformers generated by an automated semi-empirical conformational search followed by DFT optimisations. Data for the low energy conformers will be presented concurrently in the database alongside their relative energies to enable a Boltzmann-weighted summation of their contributions at the target temperature to be used in spectral predictions.


In this study, we have only considered one model chemistry ($\omega$B97X-D/def2-SVPD); however, the level of theory (e.g. density functional approximation), basis set, vibrational treatment and software package can be easily modified within the same analysis framework as new software capabilities and better benchmarking results become available. Indeed, beyond the anharmonic approach discussed above, we are also very interested to explore a hybrid approach, where harmonic calculations are performed at a very high level of theory (usually corresponding to CCSD(T) or B2PLYP calculations with larger basis sets) and the calculated frequencies and intensities are then corrected by means of GVPT2 anharmonic calculations performed at a computationally less-demanding method (e.g. hybrid functionals coupled with double-zeta basis sets) \citep{10BiPaSc.P, 14BaBiBl.P, 18BiBlPu.P}. The method has shown to provide reliable results for small to medium-sized molecules at reasonable computational times (though significantly longer than the current method), and will be considered for future work.

Our analysis has not included isotopes because the non-dominant isotope has abundances below 4.5\% in all cases except for chlorine. Nevertheless, expansion to isotopes is straightforward in {\package Gaussian} and will be considered in future work. 


Ideally, the calculated spectra should incorporate the true rotational profiles associated with the vibrational bands. The necessary band-by-band A, B and C rotational constants and dipole moments in the principal molecular axis are given within the {\package Gaussian} output file. An automated method for the generation of rotational spectra and rotational envelopes for each vibrational band from calculated rotational constants and dipole moment components will be considered in a future publication. 



\subsection{Synergies between RASCALL and CQC data}
\package{RASCALL} and our CQC approach  are symbiotic methods. \package{RASCALL} supplies preliminary data on any arbitrary molecule, providing guidance and helping to prioritize theoretical calculations. Conversely, CQC data can easily contribute to the refining, expanding, and improving the functional group data that are the primary input for the creation of \package{RASCALL} data. For example, a  major limitation of \package{RASCALL} is the reliance on good data for the prediction of the spectral behavior of different functional groups. In \package{RASCALL} 1.0 \citep{19SoPeSe.P}, these data are generated from experimental spectra and/or theoretically extrapolating from existing functional group data. Future updates to the \package{RASCALL} database can use a small number of CQC calculations to parameterise these functional group data;  specifically, infrared spectra can be computed for a representative series of molecules containing a functional group, with the average predicted vibrational frequencies and intensities extracted for the functional-group related vibrations (as identified most robustly through consideration of the vibrational eigenvectors).  In this way, a relatively small number of high  level CQC calculations can be used to parameterise \package{RASCALL}. 
Subsequently, \package{RASCALL} can predict vibrational spectra for very large molecules beyond the reach of traditional CQC methods, a key future application of this approach. 

\section{Discussion}
\label{discussions}

In this section, we discuss a few important aspects of this research, including: the diverse potential uses of these data; how the spectroscopic data work alongside the kinetic and reaction network data to enable better understanding of remote gaseous environments; and a brief discussion of the advantages and challenges of our interdisciplinary approach for biosignature detection follow-up. 



\subsection{Data Utilisation}

Our data predicts semi-quantitative spectral intensities for most of the \pmol{} studied for the first time, essential information to assess detectability in remote environments. Molecules with strong transition intensities can be far more easily detected than those with weaker transitions. For instance, on Earth, the very strong infrared absorption of \ce{CO2}, which, at over 0.041\% of the atmosphere, dominates associated spectroscopy and majorly influences global temperatures, while \ce{O2} at 21\% atmospheric concentration does not absorb infrared light due to selection rules and only has very weak (forbidden) visible transitions. The data in this paper provides sufficiently accurate intensity predictions to both rank molecular detectability and place good thresholds on the minimum observable abundance of molecules in a given environment.


Accuracy requirements for frequencies are much more demanding and certainly our CQC results, as expected, do not reach spectroscopic accuracy, unlike some molecule-specific line list approaches. Thorough error analysis is beyond the scope of this paper but is certainly a worthwhile future pursuit. For the CQC-H1 harmonic data, we estimate our errors as 38 \cm{} based on the root-mean-squared error of the scaling factor of the $\omega$B97X-D/def2-SVPD model chemistry from \cite{15KeBrMa.P} which was calculated for 119 experimental frequencies of 30 molecules, similar to other model chemistries with  hybrid functionals and augmented double-zeta basis sets. 
RASCALL errors are expected to be larger but this needs to be verified by comparison to experimentally-measured frequencies.

Despite being unsuitable for definitive molecular identification in complex gaseous mixtures such as remote atmospheres, our frequency information provides useful information for remote characterisation of gaseous environments such as planets. First, our data can be used to categorise molecules into groups that may be difficult to disambiguate with observational data at certain resolutions and spectral windows. Second, our data can help assess the difficulty of detecting a molecule or class of molecules and identify optimal spectral windows by considering the specific molecule amongst possible contaminants. For example, the major contaminant to the P-H peak prevalent in our P-molecules is the \ce{CO2} infrared absorption, which can then be closely considered as discussed above. Finally, the scope and accuracy of our data is still enough to both comprehensively build up and selectively constrain a pool of molecular candidates that may be responsible for a particular signal. 

The value of this is evident when considering the detection of phosphine on Venus and subsequent debate about whether the single observed microwave line possibly arose from a different molecule. According to the data available to the involved astronomers, the only possible contender for the signal was the nearby absorber, \ce{SO2}, which could be ruled out. However, while the available data did cover the molecules that are likely to be most abundant in that context, it was limited in coverage; the data we produce as part of this work could support a much more comprehensive investigation for similar detections in the infrared region.

Another key use of this data is the assignment of experimental spectra. For example, computational quantum chemistry calculations have been used previously to correct misassignments in P-molecule infrared spectroscopy   \citep{robertson2003ir,mcnaughton2006comment}. The CQC approach can be useful to aid molecule identification for experiments with complex molecular mixtures formed, for example, by a discharge or as reaction products. 

Finally, the generation of a large molecular dataset is worth consideration within the context of machine learning (ML). Certainly, the last few years have witnessed a delayed but definitive permeation of techniques and approaches from the latest wave of artificial-intelligence research, i.e.~deep learning, into chemistry~\citep{18budaca,20tk} and, more broadly, the physical sciences~\citep{19cacicr}. This influence has extended to the production of infrared spectra, with, for example, one study considering the hybridisation of ML and molecular-dynamics simulations~\citep{17gabema}. More recently, VPT2 calculations have been mixed with data generated by neural networks to explore anharmonic corrections to vibrational frequencies~\citep{20laabal}. However, ML is more traditionally used in processing pre-produced data, and, indeed, ML models can be trained on a variety of relations present in the dataset within this work. Certain coding packages support such an approach, for example, the Python-based {\package DeepChem}~\citep{19raeawa}, which wraps around {\package RDKit}~\citep{RDKit.P} to convert molecular \package{SMILES} codes into hashed extended-connectivity fingerprints \citep{10roha}; these breakdowns of molecular structure are useful as input feature vectors for ML models. Consequently, it is possible to efficiently learn how combinations of molecular substructures influence infrared frequencies and intensities; RASCALL is based on a similar principle derived from domain knowledge in organic chemistry that functional groups determine infrared frequencies and approximate intensities.  It is likely that ML can improve on the RASCALL dataset by providing updated functional group information extrapolated from CQC data. As a related example, \cite{20kozhca} recently explored ML for predicting infrared spectra for polycyclic aromatic hydrocarbons based on a NASA Ames dataset of more than 3,000 spectra. Given that ML benefits from the statistical power provided by big data, the high-throughput nature of our CQC results is particularly valuable in fuelling the performance of future ML models. 

\subsection{Molecules in Reaction Network Modelling}

Important species in reaction networks might be difficult to detect remotely as they have low concentrations, e.g. the important OH radical in Earth's atmosphere is mostly detected with \textit{in situ} measurements \citep{12StWhHe.P,08PiDrZa.P}. Therefore, the spectroscopic measurements need to be combined with a chemistry-based reaction network model that contains reaction rates for all molecules in the atmospheric system. For observable species, if the observed abundance is very different from the predicted abundance this could be due to incorrect model predictions, misinterpreted data, or could indicate unusual chemistry that warrants further investigation (e.g. the detection of phosphine on Venus). 

To help readers understand the strengths and limitations of existing approaches in reaction network modelling and kinetics rate predictions, we provide appendices with brief summaries of the current approaches in these fields. Appendix B provides an overview of reaction network modelling, which is important to contextualise the sources and sinks of volatile molecules. Appendix B focuses on approaches to modelling the Earth's atmosphere and references some introductory texts on the more limited reaction network modelling of exoplanets. Appendix C introduces the fundamentals of theoretical kinetics calculations, which can be used to supplement rate constants whenever they are missing from reaction networks. Popular codes for performing theoretical kinetics calculations are also referenced in Appendix C.

The application of reaction networks and kinetics modelling is considered below for the specific situation of the potential for atmospheric formation of phosphine on Venus. 

\subsubsection{Constraining Models Involving Phosphine on Venus}
\alert{The preceding sections of this paper present spectra for a wide array of \pmol{} that could feed into \ce{PH3} formation in the Venusian atmosphere. Ultimately, if \ce{PH3} were being formed from volatile \pmol{} and not through a geological process, these \pmol{} must decompose into single-phosphorus molecules that can be successively reduced to \ce{PH3}.} To understand this process and elucidate potential abiotic pathways to \ce{PH3}, we ideally want spectroscopic measurement of the Venusian atmospheric concentrations of these \ce{PH3}-precursor \pmol{}  which could greatly constraint the Venus atmospheric models. The high-throughput spectra in this paper are a first step towards these future spectroscopic measurements. 


The thick cloud layers in Venus's atmosphere around 48 -- 70 km prevent significant solar radiation from penetrating to the lower Venusian atmosphere \citep{07TiBuCr.P,20BaPeSe.P}. 
The atmospheres within and above the cloud deck, however, do receive significant solar radiation \citep{07TiBuCr.P}. Thus this middle Venusian atmosphere, like Earth, is largely driven by reactions with radical species formed through photochemistry \citep{87PrFe.P}. A photochemical network approach can therefore be used to simulate the composition of the middle atmospheres of Venus, which typically has temperatures of 200 -- 350 K \citep{20AnImTel.P}. This approach is considered \cite{20BaPeSe.P} when modelling the production and destruction of phosphine in Venus. 

However, the atmospheric processes of Venus are far less studied than for Earth and much data is missing. Even in modelling the most abundant species in the middle and lower atmospheres of Venus (\ce{SO$_x$, CO$_x$, Cl^{.}/HCl}), \cite{20BiZh.P} note that $\sim$40\% of their reaction rates used have no experimental measurements, and those that are measured are only upper limits, or for a single temperature.  This statistic reveals much is unknown about the reaction rates of core Venusian atmospheric processes, especially minor cycles like phosphorus reactions.  

\cite{20BiZh.P} highlight which rates contained in their Venus atmospheric model are of highest priority for experimental measurement or \textit{ab initio} prediction. The photochemical model of \ce{PH3} formation by \cite{20BaPeSe.P} presents similar crucial reactions that can be considered a priority in terms of: spectroscopic detection of these species (or their precursors), lab-based measurement of key reaction rates with radicals, or \textit{ab initio} calculations. These are radical-mediated reactions that could generate the direct precursors of \ce{PH3}, as shown in Scheme 1. 

\begin{scheme}[htbp!] 
\begin{align*}
\ce{P=O^{.} + H^{.} &-> P^{.} + OH^{.}} \\
\ce{P=O^{.} + H^{.} &-> PH^{.} + O^{.}} \\
\ce{HPO + H^{.} &-> PH^{.} + OH^{.}} \\
\ce{HPO + H^{.} &-> PH2^{.} + O^{.}}
\end{align*}
    \label{scheme:reduction_reactions}
    \caption{Proposed, network limiting, reactions of \pmol{} in model of  photochemical \ce{PH3} production by Bains et al.}
    \label{sch:PH3-radicals}
\end{scheme}

The spectroscopic detection and quantification of any of the intermediates shown in Scheme \ref{sch:PH3-radicals} would greatly help constrain photochemical models of \ce{PH3} formation. High quality line lists are available for \ce{PO} from ExoMol \citep{17PrJLo.P}. However, spectroscopic signatures of molecules in Scheme 1, such as the \ce{P-H} stretch, \ce{P=O} stretch and \ce{P-H} bending modes, are common to multiple \pmol{}  (see Figure \ref{fig:RA_fg-p}). Therefore a moiety-based approach to predict spectra, like RASCALL, will yield false positives, and the CQC spectra presented in this paper provide an improvement for the identification of these intermediates. In the absence of spectroscopically determined abundances of these \pmol{}, reaction network modelling must be used. 

The reactions in Scheme \ref{sch:PH3-radicals} generate immediate \ce{PH3}-precursors and are the presumed bottle-necks in the  photochemical reaction pathway. These key reactions also lack any reaction rate data. Instead, surrogate rate data from equivalent nitrogen-containing species undergoing the equivalent reaction are used \citep{20BaPeSe.P}. However, the nitrogen surrogate reaction energies differ by $\sim$50 -- 60 kJ/mol from calculated energies of the actual phosphorous compounds \citep{20BaPeSe.P,98Ch.P}. In theoretical kinetics calculations each 10 kJ/mol difference in activation energy can alter calculated reaction rates by an order of magnitude. Therefore, the use of nitrogen surrogates could lead to misestimation of rates  by several orders of magnitude, with implications on the importance of Scheme \ref{sch:PH3-radicals} reactions as network bottlenecks. 

Therefore, rate data crucially needs to be determined for the true phosphorous compounds in Scheme \ref{sch:PH3-radicals}, either experimentally or with \textit{ab initio} calculations.  The instability of \ce{HPO} and \ce{PO^{.}} limits the availability of lab-based kinetics studies \citep{20DoBlMa.P}, but their chemistry can be calculated with high-level quantum chemical methods since only 2 -- 3 atoms are involved. High-accurate composite \textit{ab intio} methods can calculate energies of these small systems to kJ/mol, or even sub-kJ/mol, accuracy \citep{16Ka.P,04TaSzCs.P,06KaRaMa.P}. After accurate calculation of the geometries and energies of these reactions, the theoretical kinetics methods outlined in Appendix B could be used to calculate reaction rates. In fact, many molecules in the atmospheric reaction networks are likely to be transient and hard to detect, so theory may provide the most viable route to good estimates of their reaction rates.

\subsection{Initial Interdisciplinary Survey Approach to Biosignature Followup}

Astrobiology and the related study of the chemistry of planetary atmospheres are such a diverse fields that no single person can be an expert on all aspects. Instead, interdisciplinary collaborative approaches are essential.

Establishing productive interdisciplinary collaborations is rewarding but challenging, and proved essential in this pilot to appreciate diverse aspects of biosignature follow-ups. We found that astronomers, geologists, origin of life researchers, experimental spectroscopists and computational spectroscopy theorists and data scientists all had significant core knowledge - sometimes trivial in their field but unknown to others and useful in combination. Identifying and refining the salient contributions of each sub-discipline - often not what was originally anticipated - and placing it within the context of this work required time and frequent communication, aided by modern technology tools.  As a concrete example, the scarcity of gaseous \pmol{}  and the relative lack of knowledge on P-molecule speciation in Earth's atmosphere was surprising to many authors. Unexpectedly, most key knowledge on gas-phase \pmol{}  came not from modern atmospheric chemistry modelling, but from origin of life research. Atmospheric chemistry expertise instead was crucial in highlighting an under-appreciated limitation of spectroscopy in remote characterisation of atmospheres; crucially important intermediates and radicals may be unobservable remotely as their reactivity makes their atmospheric lifetime extremely short and prevents atmospheric buildup to observable concentrations.

 \section{Conclusions}
 \label{conclusions}

The key new data presented in this paper is the calculated infrared spectra of 958 phosphorus-bearing molecules (\pmol{}), which represents the best available data for almost all of these molecules. These data can be useful to highlight ambiguities in molecular detection in remote atmospheres and thus prevent misassignments of spectral features while suggesting potential assignments for a given spectral signal. These data also provide sufficiently reliable intensities of different spectral features between molecules to enable evaluation of the limits of detectability for different molecules.

These data were produced with a high-throughput mostly automated methodology using computational quantum chemistry (CQC) with the $\omega$B97X-D/def2-SVPD model chemistry used to calculate harmonic frequencies and intensities (CQC-H1) for all 958 \pmol{}. Compared to the previously available \package{RASCALL} spectral data which was produced based on the frequencies of functional groups within individual molecules, these new CQC data introduce for the first time quantitatively accurate predicted intensities and frequencies data for vibrations within the fingerprint spectral region (approximately 500 - 1,450 \cm{}) that involve large molecular motions as well as improved frequency predictions for higher frequency modes through consideration of detailed chemical environmental effects. Though further improvements to our CQC-H1 approach may be obtained by performing the calculations with anharmonic methodologies like GVPT2, we identified some challenges and limitations, particularly for anharmonic prediction of modes with low force constants, and highlighted future opportunities for methodology improvements, noting that modifications of the quantum chemistry procedure are trivial to implement within our framework. We also note the recurrence of the sporadic large errors in GVPT2 $\omega$B97X-D calculations (first noted by \cite{20BaCeFu}), which seemed to affect mostly P-H stretches through for a significant number of molecules. Future work to determine an appropriate functional for anharmonic calculation is warranted as these calculations are the only data source for accurate frequencies and intensities for overtone and combination bands, which provides a more complete picture of molecular opacity and may help distinguish between some molecules.

The other key contribution of this paper is the demonstration of significant advantages with an interdisciplinary approach to follow-up of biosignature detection. Phosphine and \pmol{} are certainly of broad interest astrophysically in gas giants and as potential biosignatures, but the immediate impetus for this paper was the tentative detection of \ce{PH3} in the clouds of Venus with extraordinary high abundance \citep{20BaPeSe.P}. An important aspect of investigating this detection is to look for other gaseous \pmol{} that could be sources or sinks of phosphine in Venus and can provide insights into the possible atmospheric network that allows for the accumulation of phosphine. To identify the molecules of interest, we used two approaches; the targeted approach consolidating known or predicted chemistry to identify gas-phase \pmol{}  of particular interest for characterisation of remote planetary atmospheres, and the reaction agnostic approach which instead considered all potentially volatile stable \pmol{}  with six or fewer non-hydrogen atoms. We conclude that, given the low volatility of many \pmol{} and the relative poor understanding of gaseous phosphorus chemistry, a more reaction-agnostic comprehensive search for volatile molecules is probably the most suitable path forward for \pmol.

\section*{Conflict of Interest Statement}

The authors declare that the research was conducted in the absence of any commercial or financial relationships that could be construed as a potential conflict of interest.

\section*{Author Contributions}

LKM conceived, designed and managed the project. MNG, ESC, LSDJ, FARL, AS and JCZT collated pre-existing data.  JCZT, LKM, PK and JO developed the CQC approach. JCZT applied CQC to produce novel vibrational spectra. AS, JCZT produced the figures. JCZT, AS, ESC, FARL, JO analysed data.  CM, ER and CDT provided expert knowledge and feedback to assist with analysis. JCZT, LKM, CS, KNR, BPB, LSDJ, DJK, GGS, LS and BLT provided expert knowledge and wrote significant sections of the paper.  All authors reviewed literature and wrote, edited or provided feedback on sections of the paper. All authors reviewed and approved the final manuscript.

\section*{Funding}

KNR is supported from a grant by the Australian Research Council (DP160101792). 


\section*{Acknowledgments}
Thanks to Maria Cunningham, Maria Perez-Pe\~{n}a and Max Litherland for their enthusiastic participation in the hackathon that started this project. 

LKM would also like to thank her awesome colleagues for the writing groups and active encouragement that fast-tracked this paper to submission in the difficult 2020 year. 

This research was undertaken with the assistance of resources from the National Computational Infrastructure (NCI Australia), an NCRIS enabled capability supported by the Australian Government.

\section*{Supplemental Data}

\alert{For the purposes of review, this data has been made available at \url{https://drive.google.com/drive/folders/1FEr9eKwHmxg9EJNqW3bMhaDNQ_MxrI8g?usp=sharing}}

The supplementary data consists of:  
\begin{itemize} 
\item A read.me file explaining the full supplementary information contents;
\item A csv file listing all molecules considered with relevant information (e.g. \package{SMILES} code, boiling point);
\item A csv file with tabulated frequencies (\cm{}) and intensities (km mole$^{-1}$) including the empirical formula and {\package SMILES} code for each molecule, the mode to which the frequency and intensity belongs to, and the mode kind (i.e. fundamental, scaled fundamental, overtone or combination band);
\item A csv file containing the force constants for fundamental frequencies for the 250 molecules with GVPT2 anharmonic data available;
\item A zip file with individual folders for each molecules named by molecular formulae and \package{SMILES} codes. Within each folder there is all RASCALL, CQC-H1 and where available CQC-A1 quantum chemistry data for the molecule, along with the raw {\package Gaussian} output files and  links to all other known spectral data sources.
\end{itemize}



\section*{Data Availability Statement}
The original contributions presented in the study are included in the article/supplementary materials, further inquiries can be directed to the corresponding author.

All data produced in this paper and all \package{RASCALL} data used in this paper is available in the Supplementary Information section. ExoMol data is available at \url{exomol.com}. 

\section*{Appendix A: Anharmonic Calculations} 

\subsection*{A1: Theory Background}
\label{theory_back}

Vibrational frequency and intensity calculations in quantum chemistry are mainly performed within the so-called double harmonic approximation, where the potential energy and the dipole moment are assumed to be quadratic and linear in the normal mode coordinates, respectively \citep{06Tu.P}. This approximation is particularly appealing in quantum chemistry due to its computational ease and affordable scaling with larger systems, thus it has become a valuable tool in the interpretation of experimental vibrational spectra. However, there are two major drawbacks to the double harmonic approximation that limit its performance.  Firstly, harmonic frequency calculations tend to systematically overestimate experimental frequencies and secondly, the selection rules that govern the harmonic approximation allow only the prediction of fundamental frequencies, neglecting overtones and combination bands. Multiplicative scaling factors can be applied to the calculated frequencies to match experimental values \citep{96ScRa, 05IrRuRa, 07MeMoRa, 10AlZhZh, 12LaCaWi, 15KeBrMa, 19Ha}. However, this approach only provides improvements in the frequencies' positions without considering their respective intensities, and it has no effect on the neglected overtones and combination bands.


To obtain computationally-derived vibrational spectra in better agreement with experimental data, anharmonicity must be explicitly considered in the calculations. Variational approaches like the Vibrational Self-consistent Field (VSCF) \citep{86Bo.P, 96JuGe.P, 99ChJuBe.P, 08BoSc, 08BoCaMe.P, 13RoGe, 19PaHoJa} or Vibrational Configuration Interaction (VCI) \citep{07Ch.P, 08BoCaMe.P, 10ScLaBe} theories can be used to this end, yet they are often limited by the molecular size and memory requirements. Instead, Vibrational Second-order Perturbation Theory (VPT2; \citealt{51Ni.P}), has gained popularity in this matter as it represents a good compromise between accuracy and computational cost for medium-to-large sized molecules \citep{12BiBlCa, 17HaGeKl, 17GrCzBe, 17KiPeBe, 19BeHu}. In the context of VPT2, the vibrational Hamiltonian is divided into unperturbed ($H^{0}$) and perturbative terms ($H^{1}$ and $H^{2}$), where the former corresponds to the common harmonic Hamiltonian and the perturbative terms incorporate third and semi-diagonal fourth derivatives to the potential energy, respectively (the second perturbative term $H^{2}$ also includes a kinetic contribution from the vibrational angular momentum) \citep{19PuTaBl}. The perturbative processing of this Hamiltonian results in a handful of simple and general formulas that can be used to calculate the vibrational frequencies for fundamentals, overtones and combination bands \citep{05Ba.P, 15Bl}. In a similar fashion, equations for the calculation of vibrational intensities are also derived under the VPT2 framework by considering both mechanical and electrical (higher-order derivatives of the dipole moment function) anharmonicity \citep{06VaSt, 07VaSt, 10BaBlGu.P, 12BlBa.P, 15Bl}.

The analytical expressions for calculating both frequencies and intensities allow the simple and straightforward computation of more realistic vibrational spectra. However, a critical limitation arises when resonances or near-degenerate states appear \citep{nielsen1945vibration, 91AmHaGr}. These states in most cases lead to nearly vanishing denominators in the VPT2 working equations, thus resulting in nonphysical values for the calculated frequencies and intensities. Vibrational energies are more commonly affected by type I ($\omega_{i} \approx 2 \omega_{j}$) and II ($\omega_{i} \approx \omega_{j} + \omega_{k}$) Fermi resonances (FR), whereas vibrational intensities are plagued with both Fermi-type and the so-called Darling-Denninson resonances (DDR) ($\omega_{i} \approx \omega{j}$) \citep{40DaDe, 12BlBa.P, 15BlBiBa}. Several approaches have been proposed to deal with resonance states and here we aim to provide a general description of those most commonly used.



The first and most common approach is the so-called deperturbed VPT2 (DVPT2) method which consists on the identification and removal of resonance states from the perturbative formulation. As the resonance terms are completely disregarded from the calculations, the DVPT2 method is unable to provide a complete picture of the vibrational nature underlying systems plagued with resonance states. This drawback is overcome by the Generalised VPT2 (GVPT2) \citep{05Ba.P, 19PuBlTa} method where the identified resonance states are treated separately through variational calculations and latter reintroduced as off-diagonal terms in the computations. In both cases (DVPT2 and GVPT2) the resonant terms are identified via two consecutive tests: a frequency difference threshold ($\Delta_{\omega}$) followed by the Martin test \citep{95MaLeTa} to evaluate the deviation between the VPT2 result and a model variational calculation ($K$). Taking a different approach, the calculations can also be performed under the Degeneracy-corrected Second-order perturbation theory (DCPT2) method where all possible resonant terms are replaced by non-divergent expressions \citep{96KuTrIs}. This method allows to compute vibrational frequencies without further concerns for resonant terms in the perturbative formulation, but struggles when strong couplings between low- and high-frequency vibrations occur \citep{12BlBiBa, 15BlBiBa}. As an alternative, Blonio and co-workers developed the Hybrid-degeneracy Corrected VPT2 (HDCPT2) method that mixes both standard VPT2 and DCPT2 frameworks to calculate anharmonic vibrational frequencies \citep{12BlBiBa}. Using a transition function, the method is able to identify those states that would be better treated under a VPT2 formulation and, likewise, those with the DCPT2 method. These VPT2 variants (currently coded in the Gaussian package) are mostly based on the conventional Rayleigh-Schr\"odiner perturbation theory allowing simple algebraic equations for the anharmonic frequencies and intensities. However, alternative significant work dealing with resonance states in VPT2 has also been performed recently, considering Van Vleck perturbation theory instead \citep{14KrIsSt, 14RoPo}. 

Despite the advantages provided by the aforementioned VPT2 flavours, there are some considerations that are worth mentioning. Though default values have been defined for the appropriate identification of resonant states in the DVPT2 and GVTP2 methods, in most cases, it is recommended to assign these values based on the specific molecular system under study. This limitation clearly hinders any high-throughput calculation of vibrational spectra as defining appropriate thresholds becomes impractical when assessing hundreds of molecules at once. Instead, one could make use of the HDCPT2 method that performs similarly to GVPT2, with the advantage of a threshold-free formulation. However, the current version of HDCPT2  only allows the calculation of vibrational frequencies and the extension to vibrational intensities is still under study \citep{12BlBiBa}. We thus use GVPT2 in our calculations. 

Finally, it is important to note that, due to their perturbative nature, though all VPT2 approaches generally perform well for semi-rigid molecules, there are substantial errors when dealing with large-amplitude vibrations, torsion and inversion modes, in the presence of double-well potentials and when considering floppy molecules \citep{12BaBiBl, 16BlBaBi.P, 17GrCzBe, 19PuTaBl, 19PuBlTa}.

\subsection*{A2: Further Computational Details}
  For the CQC-A1 approach, the GVPT2 method was used as it allows a general treatment of resonance states affecting both frequencies and intensities. These resonant states were identified using the following default thresholds:

\begin{center}
    $\Delta_{\omega}^{1-2}$ = 200 \cm{}; $K^{1-2}$ = 1 \cm{} 
    
    $\Delta_{\omega}^{2-2}$ = 100 \cm{}; $K^{2-2}$ = 10 \cm{}
    
    $\Delta_{\omega}^{1-1}$ = 100 \cm{}; $K^{1-1}$ = 10 \cm{}
\end{center}

Where the $1-2$ superscript corresponds to Fermi-type resonances and both $2-2$ and $1-1$ represent Darling-Denninson resonances. The cubic and semidiagonal quartic derivatives of the potential were obtained by numerical differentiation of the analytic second derivatives, with the default 0.01 {\AA} step.

\subsection*{A3: Discussion of Anomolous Anharmonic Results}

\begin{figure}[htbp!]
    \centering
    \includegraphics[width=0.48\textwidth]{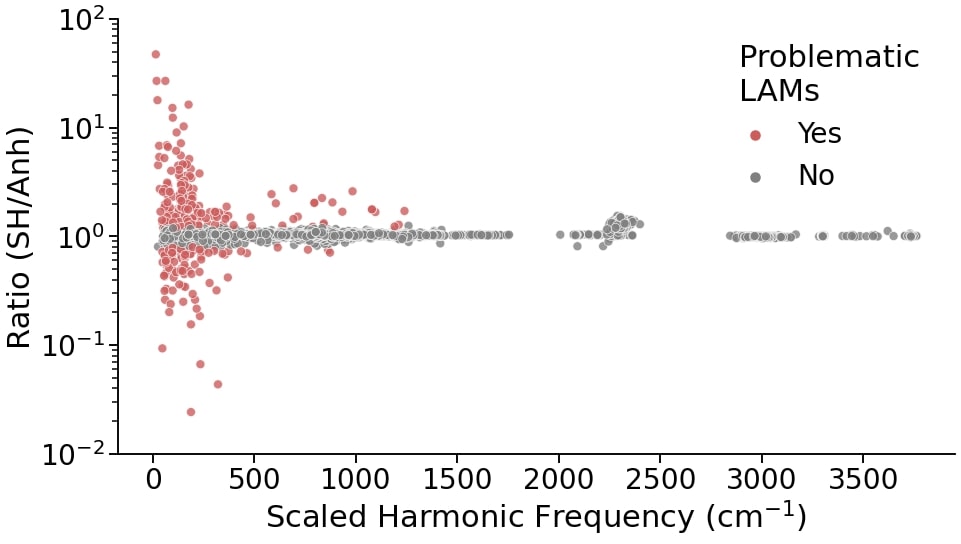}
    \includegraphics[width=0.48\textwidth]{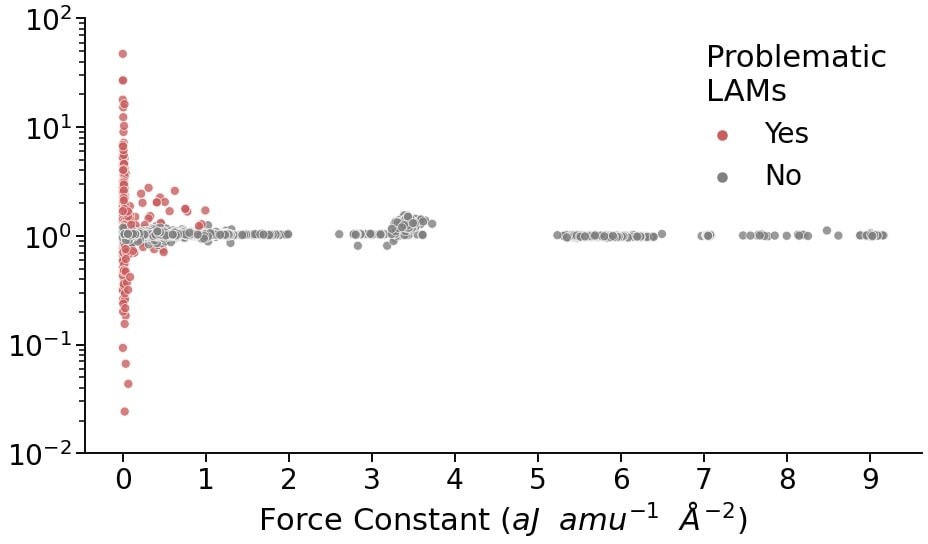}
    \caption{Ratio between the scaled harmonic (SH) and GVPT2 anharmonic (Anh) frequencies as a function of the scaled harmonic frequencies (left panel) and force constants (right panel) for 250 molecules, highlighting problematic modes due to large amplitude vibrations (LAMs) in our data. Both y axes are given in logarithmic scale.}
    \label{fig:ratio_fc}
\end{figure}

Figure \ref{fig:ratio_fc} shows the ratio between the scaled harmonic (SH) and anharmonic fundamental frequencies (Anh) as a function of frequency (left panel) and force constant (right panel). Generally, this ratio is very close to 1, i.e. the harmonic and anharmonic calculations have very similar predictions. However, there are two cases where the calculations differ substantially: (1) many low-frequency transitions with small force constants have very large differences (up to a factor of 50) between the harmonic and anharmonic frequency predictions and (2) the P-H stretch frequencies are in many molecules 200-700 \cm{} or more higher in the anharmonic compared to the harmonic calculations. In summary, the first issue is well-known for large amplitude vibrations (LAMs) and is an inherent limitation of perturbative approaches, while the second issue is far more concerning and can be traced back to deficiencies in the density functional. 

The first issue, i.e. large scaled harmonic-anharmonic frequency (SH/Anh) differences for transitions with small force constants, is well known. The small force constants are characteristic of large amplitude vibration (LAMs) which are motions that occur along very flat areas of the potential energy surface.  GVPT2 anharmonic frequencies for large amplitude vibrations (LAMs) can greatly differ from the scaled harmonic ones, due to the unsatisfactory performance of perturbation theory in these cases \citep{12BaBiBl, 16BlBaBi.P, 17GrCzBe, 19PuTaBl, 19PuBlTa}. Under the assumption that perturbation theory fails when the correction is too large, we have chosen to flag anharmonic calculations as problematic due to LAM all frequencies with SH/Anh ratio $<$ 0.8 and SH/Anh ratio $>$ 1.2 that also have a small force constant $<$ 1 $aJ$ $amu^{-1}$ $\AA^{-2}$.

\begin{figure}[htbp!]
    \centering
    \includegraphics[width=0.95\textwidth]{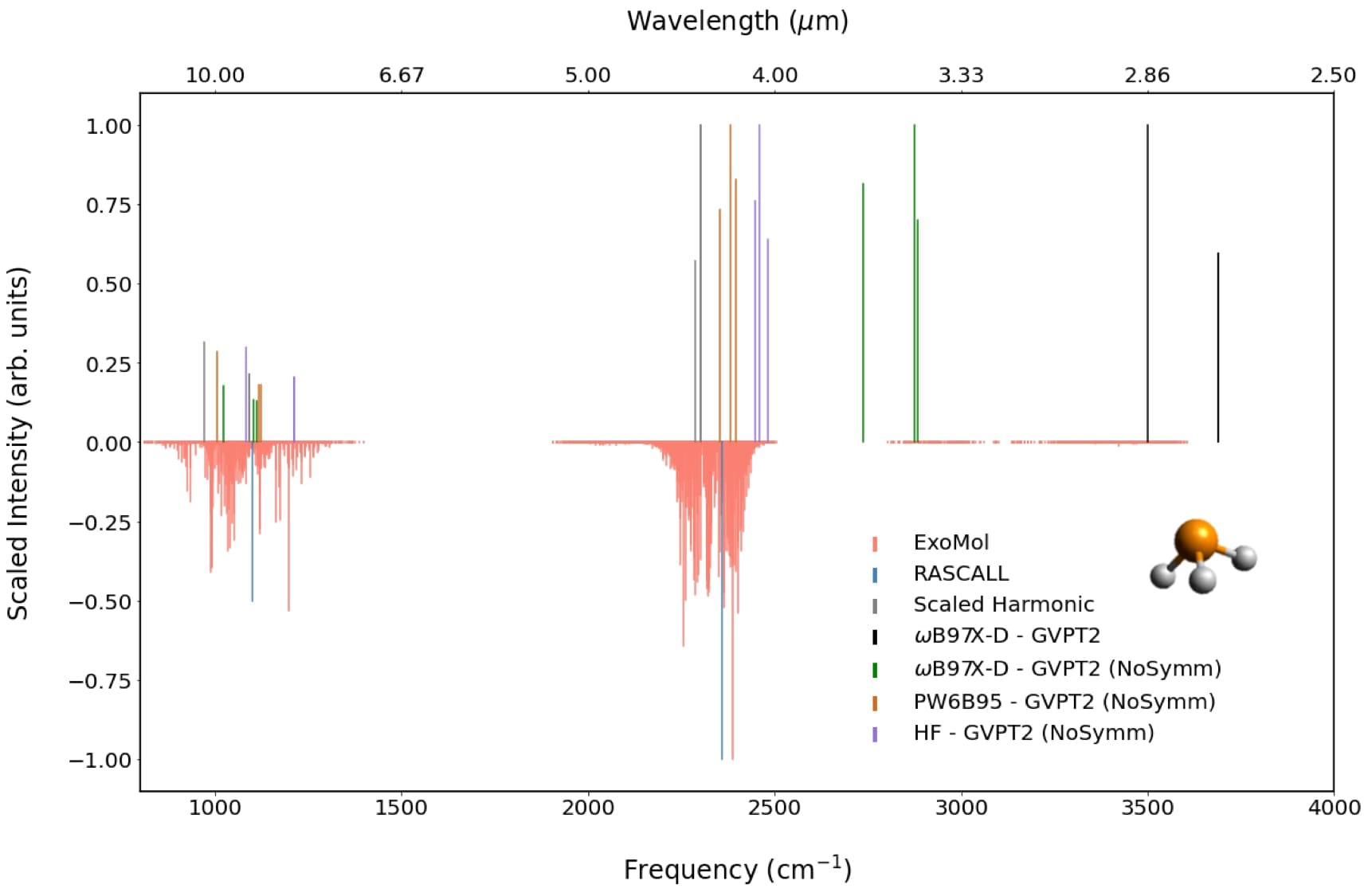}
    \caption{Comparison of the ExoMol, RASCALL, scaled harmonic and GVPT2 anharmonic frequencies for phosphine. A demonstration of the limitations present in the GVPT2 procedure for highly symmetric systems.}
    \label{fig:PH3}
\end{figure}

The second issue, large SH/Anh ratios for P-H stretches, is more unusual. The prototypical example is \ce{PH3}, with results shown in Figure \ref{fig:PH3} for the harmonic, anharmonic, \package{RASCALL} and reference ExoMol spectra for \ce{PH3}. For the P-H stretch signal (near 2,360 \cm{}), the scaled harmonic and anharmonic calculations differ by over 1,000 \cm{}, with the scaled harmonic calculations seen as reliable by comparison to the ExoMol and \package{RASCALL} data sources. The highly symmetric nature of \ce{PH3} causes intrinsic degeneracies that results in unreliable vibrational frequencies and intensities. Some improvement can be obtained by formally lowering the symmetry in the calculations (NoSymm option in Gaussian), as seen in Figure \ref{fig:PH3} comparing the results from the $\omega$B97X-D - GVPT2 (black) and $\omega$B97X-D - GVPT2 - NoSymm (green) calculations. However, unacceptably large errors still occur of 100s of \cm{}, and the harmonic calculations are far superior (compared to experimental data). Further, similar very large P-H anharmonic stretch frequencies were found in many unsymmetric molecules; for example, the \ce{CH5OP} isomer with SMILES code OCP, exhibited differences of 405 and 467 \cm{} between the scaled harmonic and anharmonic P-H stretch frequencies as calculated by $\omega$B97X-D/def2-SVPD. Therefore, use of NoSymm is helpful but not sufficient for resolving the issue of large SH/Anh ratios for P-H stretches.

We considered neglected resonances as a possible cause for large SH/Anh ratios for P-H stretches, but our tests showed that deuterating one of the hydrogens in the \ce{CH5OP} isomer (which should alter the resonance patterns) did not remove these errors.

Recent results from \citep{20BaCeFu} (published subsequent to our calculations) suggested that the density functional might be the issue; this benchmark study covered ten small molecules and found that the $\omega$B97X-D functional can yield unstable anharmonic vibrational frequencies, potentially due to an inappropriate calculation of higher-order derivatives of the potential energy. Our \ce{CH5OP} isomer output files showed warnings for large cubic and quartic force constants. We confirmed the culpability of the functional through tests of HF/def2-SVPD and PW6B95/def2-SVPD calculations for the \ce{PH3} and \ce{CH5OP} isomer that did not show the very large scaled harmonic-anharmonic shifts observed with $\omega$B97X-D. Therefore, we conclude that a different choice of functional is required for reliable anharmonic calculations, despite the very good performance of $\omega$B97X-D for general chemistry (e.g. \cite{goerigk2019trip, 20ZaMc.P}). We defer detailed consideration of alternative functionals with anharmonic methods to a future publications in order to enable detailed comparison to experiment for a large number of molecules.

However, new density functionals are unlikely to enable the accurate anharmonic treatment of large amplitude modes, for which the harmonic approximation is not a sufficiently accurate approximation for perturbative treatments to be reliable; more expensive variational approaches like VCI are likely to be required. In  automated high-throughput approach, the most practical solution is probably the identification of problematic transitions through flags, as described above. We will undertake future investigations that explicitly compare predicted harmonic and anharmonic computed frequencies against experimental data for a large number of molecules. This will enable us to improve this flagging process as well as develop methods to automatically identify likely problematic molecules.

\section*{Appendix B: Reaction Network Modelling}
Significant work has gone into detailed kinetic models of the chemical reaction networks in the Earth's atmosphere. GEOS-Chem \citep{GEOS-Chem} and the Master Chemical Mechanism (MCM) \citep{MCM} are two large scale chemistry focused models, with the MCM aiming to be a `near-explicit' mechanism: modelling 6,500+ chemical species and 17,000+ reactions. This reaction network is developed from the explicit representation of the degradation of 143 volatile organic compounds (VOCs), with pre-defined rules of chemical reactivity following an initiation reaction (e.g. oxidation, ozonolysis, photolysis), generating the vast number of reactions that need to be represented \citep{03Sa.P}. Not all species represented in the atmospheric models have well-known behaviour, and these less well understood species are propagated through the reaction network until they form end-products whose chemistry is known (e.g. CO, ROH species). 

The aim of making atmospheric models comprehensive is complicated by a combinatorial explosion of possible species and minor reaction channels. To remedy the overabundance of VOCs to be modelled, a process known as `lumping' is used \citep{98WaGeLi.P,04WhToPi.P}, where only species with branching fractions $>$5\% are explicitly represented \citep{97JeSaPi.P}, and a generic intermediate is used to represent all molecules that ultimately form similar products. A fundamental challenge of modelling the Earth's atmospheric chemistry is therefore the reduction of model complexity, so the models are computationally tractable and can be used for simulation. 

On Earth, field campaigns and lab-based measurements can more easily acquire relevant experimental data, in contrast to the data needed for exoplanet atmospheres. Nevertheless, due to the diversity of VOCs known and predicted, Earth's atmospheric models are increasingly relying on theoretical methods to fill out missing experimental data. A review of how theory, and its interplay with experiment, has advanced our understanding of Earth's atmospheric reactions networks is provided in \citep{12VeFr.P}, with practical examples given in \citep{15VeGlPi.P}. In Appendix C, various theoretical kinetics methods for predicting reaction rates are outlined briefly.


Exoplanetary atmospheres will be far more diverse, with hot gas giants a particular current focus as these will be the first exoplanet atmospheres to be characterised as they are easiest to observe. The reaction networks are generally more limited, but are appropriate for a wider range of physical conditions and temperatures.  \cite{17CaKa.Atmosphere,17Heng.ExoAtm} provide good introductions to exoplanetary atmosphere modelling.  



\section*{Appendix C: Kinetic Rates}\label{sec:kinetic_rates}
Prediction of molecular reaction rates is usually more challenging both experimentally and computationally than thermochemistry, with accuracies within an order of magnitude for rates generally considered very good. 
Molecular reaction rates can be determined using transition state theory (TST) \citep{00Pe.P}. The foundations of TST was the insight by \cite{35Ey.P} that the transition state represents a dividing surface that minimises the reactive flux between reactants and products. The transition state is a first order saddle point: the point of maximum energy (i.e. bond strain) along the minimum energy reaction pathway. Standard TST rates can be calculated for bimolecular reactions from the activation energy, and partition functions of the TS and the reactants. Unimolecular reaction rates can be calculated by Rice-Ramsperger-Kassel-Marcus (RRKM) theory, where the reaction rate at a given energy is the ratio of the density of (vibrational) states at the minimum energy well and at the TS \citep{73Fo.P}. The vibrational states are usually treated as coupled harmonic oscillators with fast intramolecular vibrational energy redistribution between them. 

Where a reaction is barrierless along the minimum energy pathway, i.e. does not have an activation energy, modifications to TST are required. An example of barrierless reactions are radical-radical recombination reactions, which combine with no barrier, or conversely, dissociation of a closed-shell molecule into two radicals. These barrierless reactions are often important to photochemistry and atmospheric chemistry. In barrierless reactions, a variational approach (VTST) is used to determine a dividing surface along the reaction coordinate that minimises the reactive flux by maximising the Gibbs free energy through consideration of the entropy changes of the reactants  \citep{17BaTr.P}. For reactions with a ``loose" transition state, the variable reaction coordinate (VRC-TST) approach can be used to determine the dividing surface more flexibly, through use of optimised pivot points between the reacting fragments  \citep{03GeKl.P}. Several codes are available with different methodologies for calculating barrierless reaction rate constants, including: MultiWell (VTST) \citep{01Ba.P}, Polyrate (VRC-TST) \citep{17ZhBaMe.P}, and MESMER where the inverse Laplace transform approach is used when high pressure limit data are available \citep{12GlLiMo.P}. 

Theoretical rate constants can often deviate by one or two orders of magnitude from experiment, but provide mechanistic insight and are generally more suitable to model a particular reaction than a ``surrogate" experimental reaction rate from an analogous system. Theoretical kinetics therefore provides an important pathway to supplement reaction networks with estimates of reaction rates whenever experimental data are missing or difficult to obtain.

\bibliography{references}
\bibliographystyle{frontiers2}

\end{document}